\documentclass{aa}
\def\msol{{M}_{\odot}}
\def\lsol{{L}_{\odot}}
\input psfig
\begin{document}

   \thesaurus{03     
              ( 11.03.4; 
                11.09.3; 
                12.03.3; 
                11.01.2; 
                13.25.2)} 

\title{X-ray observations of the rich cluster CL 0939+4713 and
                discovery of the strongly variable source
                RXJ0943.0+4701} 


   \author{	Sabine Schindler$^{1,2,3}$, 
		Paola Belloni$^4$, 
                Yasushi Ikebe$^1$,
		Makoto Hattori$^5$,
   		Joachim Wambsganss$^6$,
		Yasuo Tanaka$^{1,7}$
	}

   \institute{
               MPI f\"ur extraterrestrische Physik,
               Giessenbachstr.,
               85748 Garching, 
               Germany; 
               e-mail: {\tt sas@staru1.livjm.ac.uk} 
\and
       	       MPI f\"ur Astrophysik,
               Karl-Schwarzschild-Str. 1,
               85748 Garching, 
               Germany 
\and
               Astrophysics Research Institute, 
               Liverpool John--Moores University, 
               Byrom Street, 
               Liverpool L3 3AF, 
               United Kingdom
\and
       	       Universit\"atssternwarte M\"unchen,
	       Scheinerstr. 1,
               81689 M\"unchen, 
               Germany 
\and
               Astronomical Institute,
               T\^{o}hoku University, 
               Aoba Aramaki,
               Sendai 980, 
               Japan: 
               e-mail: {\tt hattori@astroa.astr.tohoku.ac.jp}
\and
       	       Astrophysikalisches Institut Potsdam,
               An der Sternwarte 16,
               14482 Potsdam,
               Germany;
               e-mail: {\tt jwambsganss@aip.de}
\and
       	       Institute of Space and Astronautical Science, 
               Yoshinodai 3-1-1, Sagamihara,
               Kanagawa 229, Japan
              }
   \date{}
\authorrunning {Sabine Schindler et al.}
\titlerunning {X-ray observations of the rich 
cluster CL 0939+4713}
   \maketitle

   \begin{abstract}

Recent X-ray observations of the rich galaxy cluster CL 0939+4713 
(or Abell 851) at $z = 0.41$ with the ROSAT/HRI as well as  with 
the ASCA/GIS and ASCA\-/SIS instruments are presented and analysed. 
With the high resolution imaging data (ROSAT/HRI) we  confirm and 
extend the earlier impression that the cluster has a lot of 
substructure. 
Two  maxima of the cluster emission are obvious in the images, 
each of them shows even some internal structure. One of the subclusters 
can be nicely modeled with an elliptical model. 
For the total luminosity of the cluster in the ROSAT band we obtain 
	$L_{X, 0.1-2.4keV} = 6.4_{-0.3}^{+0.7}\times 10^{44}$ erg/s, 
for the bolometric luminosity 
	$L_{X,  bol} = 1.6_{-0.3}^{+0.5}\times 10^{45}$ erg/s. 
We perform spectral fits for the two ASCA instruments and 
for the ROSAT/PSPC simultaneously. 
The most reliable numbers for the temperature and metallicity for the
intracluster gas are 
	$T = 7.6^{+2.8}_{-1.6}$ keV
and
	$m = 0.22^{+0.24}_{-0.22} m_{\odot}$, 
respectively. We find a relatively
small total mass, a small gas mass ratio and a
small iron mass to light ratio. These numbers together with relatively
low luminosity for such an optically rich cluster and the pronounced
substructure confirm that CL 0939+4713 is a young cluster still far
away from a virialised state.

In the same ROSAT/HRI image 
the X-ray emission of a background quasar ($z_Q \approx 2$)  
can clearly be identified. 
With  a luminosity of 
	$L_{X, 0.1-2.4keV} \approx  1.4\times 10^{45}$ erg/s 
(which is necessarily affected by gravitational lensing)
it belongs to the X-ray brightest quasars.
A striking difference  between the ROSAT/PSPC and the ROSAT/HRI
observations (taken almost five years apart) is the   
fainting of one X-ray source (RXJ0943.0+4701) by at least a factor of ten. 
We try to identify this source from deep optical data. Our best 
candidate is a blue compact object, possibly an AGN.

      \keywords{Galaxies: clusters: individual: CL 0939+4713 --
                Galaxies: clusters: individual: Abell 851 --
                intergalactic medium --
                Cosmology: observations --
                Galaxies: active --
                X-rays: galaxies
               }
   \end{abstract}

%
%

\section{Introduction}

One of the foremost goals of modern cosmology is the understanding of
the formation and evolution of structure. Clusters of galaxies provide 
an excellent probe because they represent the largest gravitationally
bound objects in the universe and can be identified over a large 
range of redshifts. Studying the X-ray properties of distant clusters
and comparing them with the properties of nearby clusters gives 
important insight into the evolution of galaxy clusters and provides
a strong test for cosmological models.

The galaxy cluster presented in this paper is the  
rich cluster CL 0939+4713 at redshift $z = 0.41$. It is very
well studied optically (see e.g.  
	Dressler \& Gunn  1983; 
	Dressler \& Gunn 1992;
	Dressler et al.  1993; 
	Belloni et al. 1995; 
	Belloni \& R\"oser  1996). 
Almost 200 (spectroscopic and photometric) redshifts are known.
The cluster
is extremely rich, and it contains a high fraction of blue and E+A galaxies.
A background quasar at redshift two is seen through the cluster.
CL 0939+4713 was subject of a weak lensing analysis (Seitz et al. 1996),
and very recently Trager et al. (1997) showed that there are
a few redshift four galaxies seen through the cluster.

A ROSAT/PSPC observation of this cluster was 
analy\-sed by Schindler \& Wambsganss (1996).
It was shown that the X-ray luminosity is
relatively low for such a rich cluster. 
Meanwhile new X-ray data are available from the high resolution
ROSAT/HRI detector as well as improved spectral information 
from the ASCA/GIS and SIS instruments.

Here we present and analyse this new X-ray data. 
We investigate the 
X-ray morphology of the  cluster (Sect. \ref{subsec-morph}) from
the new ROSAT data and show that the cluster contains clear 
substructure. In Sect. \ref{subsec-spec} we present a spectral analysis
from both ASCA instruments and from the ROSAT/PSPC data. 
The X-ray luminosity is determined in Sect. \ref{subsec-lum}. 
In Sect. \ref{subsec-mass} we estimate the gas mass, the iron mass
and the total mass.
We  look in some detail into the variable X-ray source RXJ0943.0+4701 
whose flux weakened by more than a factor of ten over the
five years between the two ROSAT observations
(Sect. \ref{subsec-m3}). The attempt to identify an optical 
counterpart for RXJ0943.0\-+4701 is described in Sect. \ref{subsec-opt}.
Finally we identify and measure the X-ray properties of
the quasar in the field (Sect. \ref{subsec-quasar}).
In the final Chap. \ref{sec-dis} we discuss
and summarize  the X-ray and optical properties of this
unusual galaxy cluster. 

Throughout this paper we use $H_0 = 50$ km/s/Mpc.

\section{Observations}

For the analysis we use optical and X-ray observations -- taken with
ROSAT (Tr\"umper 1983) and ASCA (Tanaka et al. 1994) -- obtained within
a period of five years. The X-ray observations consist of three data
sets taken with different instruments:
\begin{itemize}
\item 14.4 ksec with ROSAT/PSPC on November 15$^{th}$ to 20$^{th}$, 1991 
\item 39 ksec with ASCA       on April 12$^{th}$ and 13$^{th}$, 1995 
\item 45.6 ksec with ROSAT/HRI  on October 16$^{th}$ to 20$^{th}$, 1996 
\end{itemize}
While the ROSAT observations give us information on the spatial
distribution of the X-ray emission, the ASCA data -- and to some degree
also the ROSAT/PSPC data -- provide spectral resolution. The ASCA
analysis uses both types of detectors -- the Solid-state Imaging
Spectrometers (SIS) and the Gas Imaging Spectrometers (GIS). The SISs
were operated in 2CCD Faint mode,  
while the GISs were operated in PH mode 
(see http://asca.gsfc.nasa.gov/docs\-/asca/abc/abc.html). The ASCA data are 
selected according to the following criteria:
elevation higher than 20 degrees and bright Earth angle larger than 
30 degrees.

 The optical observations have been all obtained in November 1992
 at the prime focus of the 3.5 m telescope on Calar Alto (Spain)
 and are listed in Table 1 of Belloni et al. (1995).
 A series of narrow-band filters had been used
 (FWHM $ \simeq$ 90--200 \AA) covering the range from 3800 to
 9200 \AA, and broad band B, R, I filters to obtain low
 resolution spectra for all galaxies brighter than 
 R=22.5 mag in a 7 arcmin x 7 arcmin
 cluster field. The filters  were selected to examine as
 accurately as possible the spectral energy distribution (SED) 
 of ellipticals and E+A
 galaxies (Dressler \& Gunn 1983) at the clusters' redshift.  
 Therefore, they allow us to estimate the redshift of these galaxies
 with an accuracy $\sigma_{z}=0.010$ (Thimm et al. 1994)
 and with an accuracy of about $\sigma_{z}=0.030$ for
 cluster members of other morphological types
 and foreground/background galaxies.

\subsection{X-ray morphology of CL 0939+4713} \label{subsec-morph}

The ROSAT/PSPC data of cluster CL 0939+4713 were analysed and discussed
in Schindler \& Wambsganss (1996). Here we concentrate on the higher
resolution and much longer HRI observation and compare the two data
sets.

The ROSAT/HRI image of CL 0939+4713 obtained in October 1996 is
shown in Fig. 1. The cluster looks very amorphous.
It it obvious that there is a lot of substructure in
this cluster. 
The cluster emission consists mainly of two maxima labeled M1 and M2
(see Fig. 1.).
Both of them seem not spherically
symmetric but show some internal structure. The positions of these
maxima are: 
\begin{itemize}
\item M1 at $9^h42^m58.2^s$, $46^{\circ}58'52''$ (J2000),  
	with a maximum surface brightness of 
	$9.2\times 10^{-3}$ erg/arcmin$^2$/s, and 
\item M2 at $9^h43^m04.4^s$, $46^{\circ}59'48''$ (J2000) 
	with a maximum surface brightness of 
	$10.2\times 10^{-3}$ erg/arcmin$^2$/s.
\end{itemize}
These values are obtained from an image smoothed
with a Gaussian filter of $\sigma=10$arcsec. The central part of the
ROSAT/HRI image looks very similar to the smoothed light distribution
of all identified E/S0 galaxies (Seitz et al. 1996, Fig. 9b) as
observed by Dressler et al. (1994).

The high resolution of the HRI allows for the first
time to resolve the emission of the quasar at z=2.055 (Dressler et
al. 1993). It is located left of the left maximum and marked
with a plus sign in Fig. 1.

The most striking difference between this HRI observation  and
the earlier PSPC image is the complete absence of
the emission in the north which is marked as M3 (RXJ0943.0+4701) in the
PSPC observation (see Schindler \& Wambsganss 1996).
This will be investigated in detail in Section \ref{subsec-m3}.
The bright point source (P1) in the left part of Fig. 1  
was detected already with the PSPC. It is very likely not
related to the cluster.

\begin{figure*}
\psfig{figure=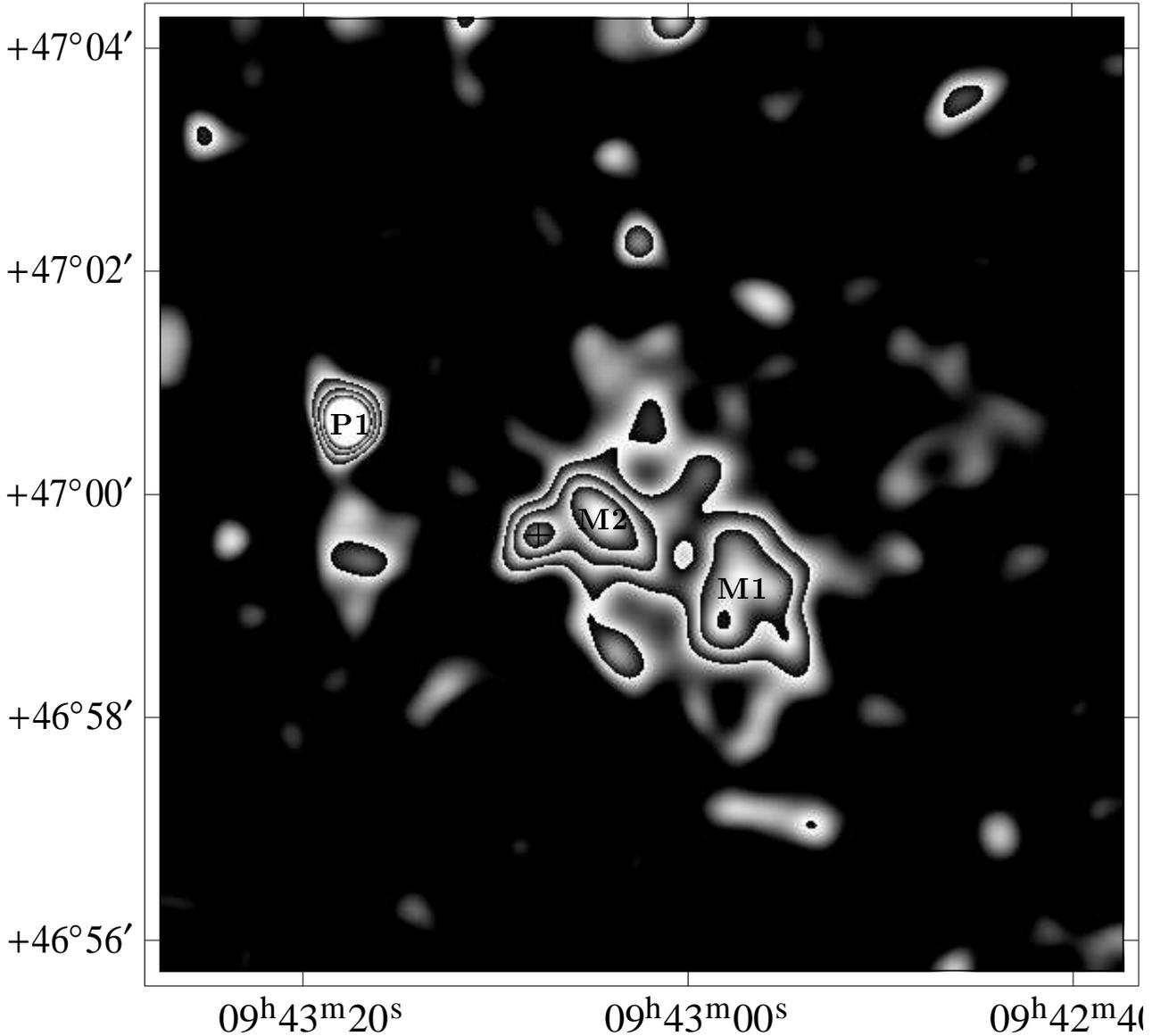,width=17cm,clip=} \label{fig-hri}
\caption[]{ROSAT/HRI image of the cluster CL 0939+4713. The image is
smoothed with a Gaussian filter of $\sigma=10$arcseconds. The cluster
shows very pronounced substructure. The two main maxima are marked
with M1 and M2.
The extension left of the second
maximum is caused by a quasar at $z=2.055$. Its exact position
is indicated by the plus sign. The bright point source (P1) is
very likely not related to the cluster.
}
\end{figure*}

As the cluster is clearly not spherically symmetric we do not attempt
to fit a single $\beta$ model (Cavaliere \& Fusco-Femiano 1976; Jones \&
Forman 1984)

$$S(r) = 
S_0 \left( 1 + {\left({r \over r_c}\right)}^2\right)^{-3\beta + 1/2},
   \eqno(1)
$$
to the cluster as a whole, but to the two subclusters separately ($S(r)$ is
the surface brightness at distance r, 
$S_0$ is the central surface brightness, $r_c$ is the core
radius, and $\beta$ is the slope parameter).
The fits are
centred on the respective maximum, a sector of $90^\circ$ in the
direction of the other
subcluster and the quasar emission being excluded. The profiles and the
fits are shown in Fig. 2, the fit parameters are listed in Table 1.
Both subclusters show very small values for $\beta$ and for the core
radius. 

\begin{figure*} \label{fig-prof}
\begin{tabular}{ll} 
\psfig{figure=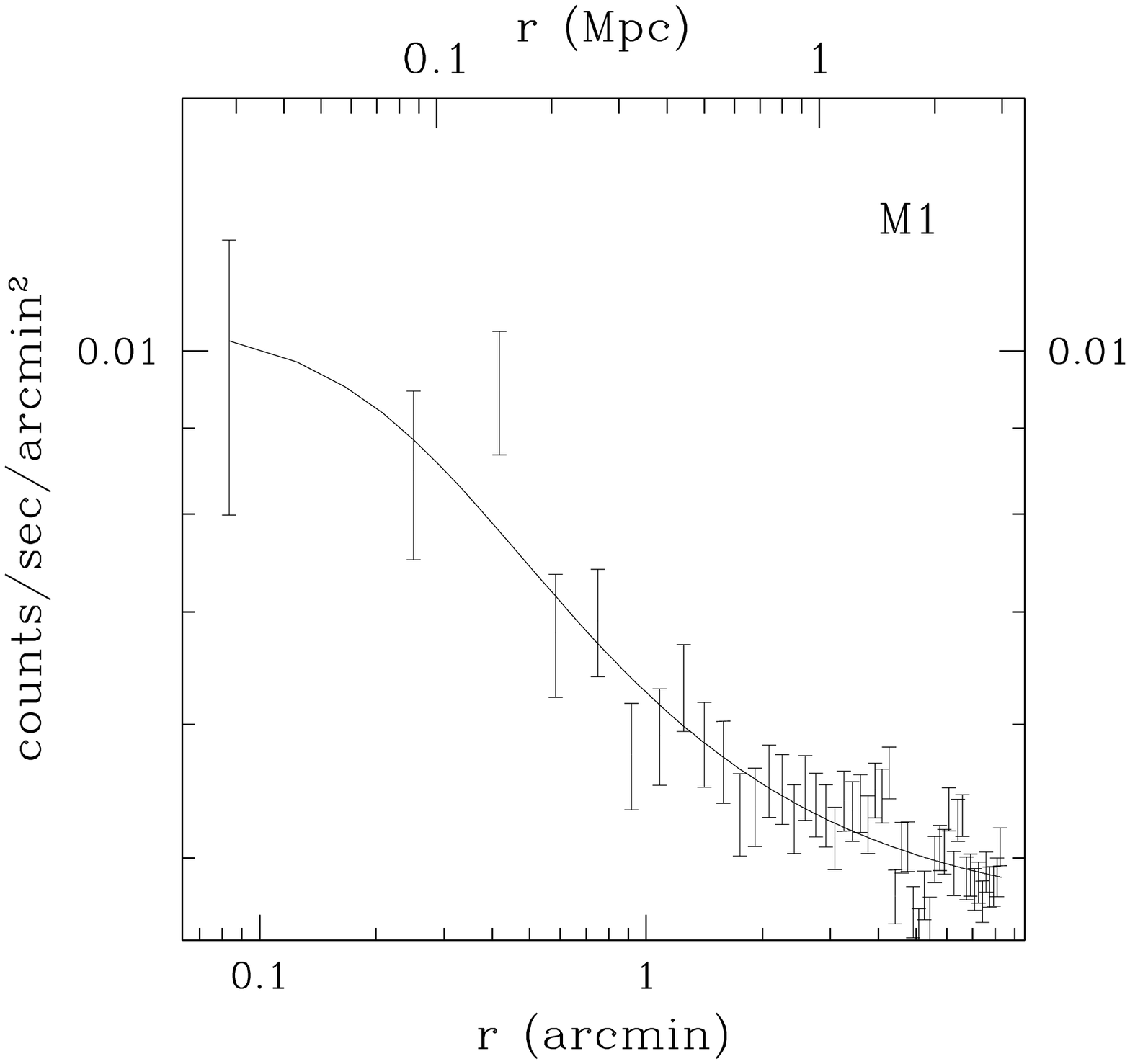,width=9.5cm,clip=} &
\psfig{figure=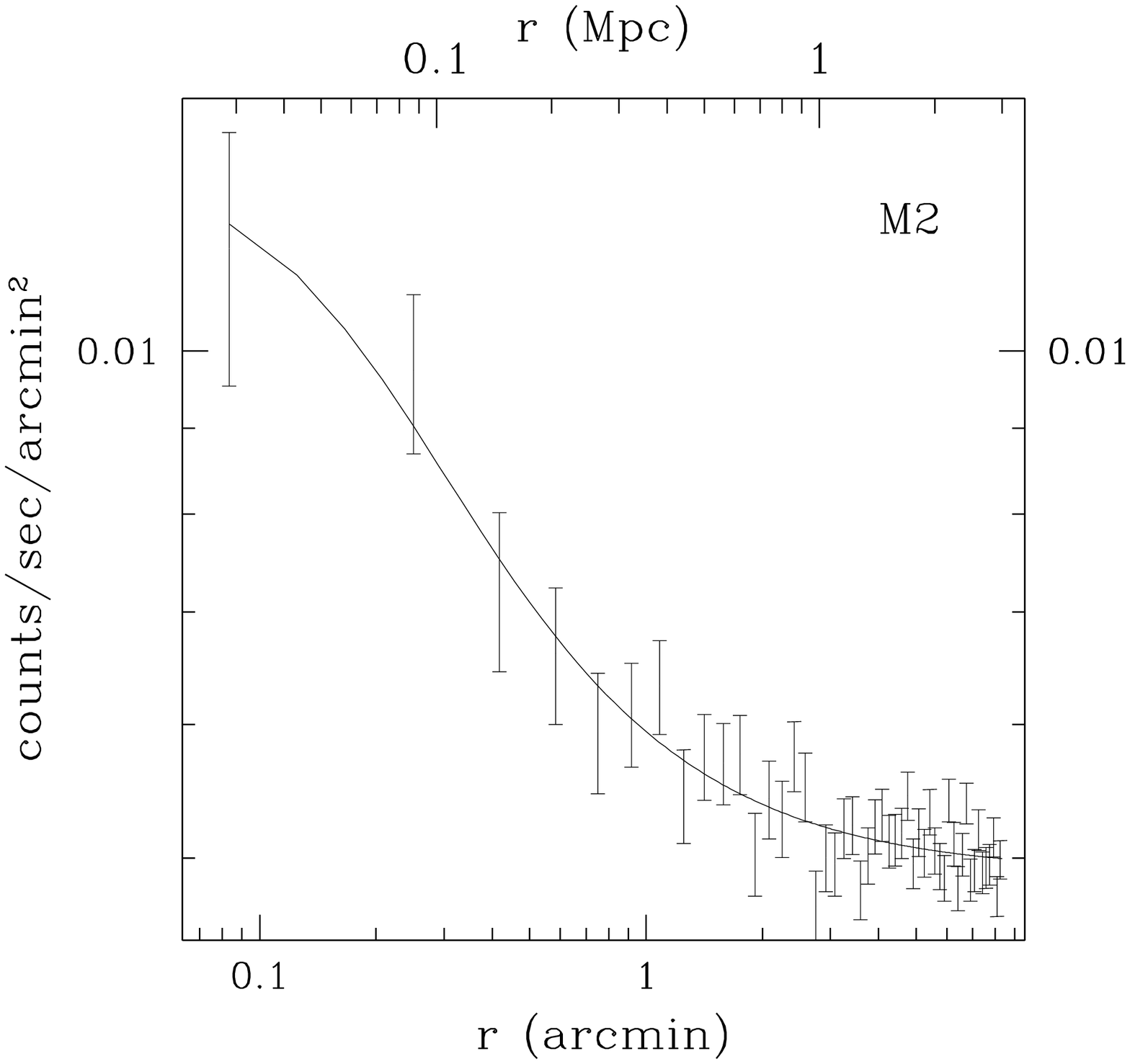,width=7.9cm,clip=} \\
\end{tabular}
\caption[]{Surface brightness profiles around the maxima M1 and M2
with a spherically symmetric $\beta$-model fit. The fit parameters are
given in Table 1.
}
\end{figure*}

For a quantitative analysis of the substructure we apply the method
by Neumann \& B\"ohringer (1997): we fit an elliptical $\beta$ model
to the HRI data. 
Since the cluster has a bimodal shape an elliptical fit cannot converge.
Therefore we have to exclude one of the maxima for fit. 
Both maxima have about the same number of counts within a radius of 40
arcseconds, but M1 is less peaked  and its shape  corresponds less
to an elliptical $\beta$ profile. 
Therefore we exclude M1 and centre the fit on M2. 
The quasar emission is excluded for the fit as well. 
The resulting fit parameters are given in Table 1. They correspond
well with the parameters of the spherically symmetrical fit,
hence  we can be sure that the elliptical fit converged reasonably well. 
Figure 3a shows the elliptical model. 

%
%
%
%
\begin{table*}[htbp]
\begin{center}
\begin{tabular}{|c|c|c|c|c|c|c|}    
\hline
& & & & & &\\
Type of fit&centre&S$_{\rm 0}$ in cts/s/arcmin$^2$ &r$_{\rm c}$ in
arcsec&r$_{\rm c}$ in kpc  &$\beta$&PA  (N over E) \\
& & & & & &\\
\hline
& & & & & &\\
spherical fit  &M1&$5.8\times10^{-3}$ &15& 95         &0.31& - \\
spherical fit  &M2&$7.9\times10^{-3}$ &11& 67         &0.36& - \\
elliptical fit &M2&$8.3\times10^{-3}$ &12(major axis) &79(major axis)&0.36&133$^\circ$\\
               &           &                   &9(minor axis) &56(minor axis)&&\\
\hline
\end{tabular}
\end{center}
\caption{Fit parameters for the spherically symmetric and elliptical
$\beta$ model fits to the two subcluster: 
$S_0$ is the central surface brightness,
$r_c$ is the core radius (for the elliptical fit we give two numbers
for the major and the minor axes), $\beta$ is the slope parameter, and
PA is the position angle (only for the elliptical fit).}
\end{table*}

Subsequently, we use the elliptical model and subtract it from the 
original image. 
The residuals are shown in Fig. 3b. 
The emission of and around M2 has almost disappeared which indicates
that the elliptical fit is a good model.

In this residual map the maximum M1 is clearly visible, it
has a significance of more than 4$\sigma$.  
It is quite obvious from the morphology that M1 is not relaxed  
but has some internal structure. 
There is also some diffuse emission surrounding M1, 
but with lower significance. The emission of the quasar is very
clearly visible.
The other emission around the cluster has only significances 
between 2 and 3$\sigma$, i.e. it
is not clear whether each of these features is indicating
more substructure is real or whether some  of them are
just statistical fluctuations.

\begin{figure} \label{fig-ell-mod}
\begin{tabular}{c}   
\psfig{figure=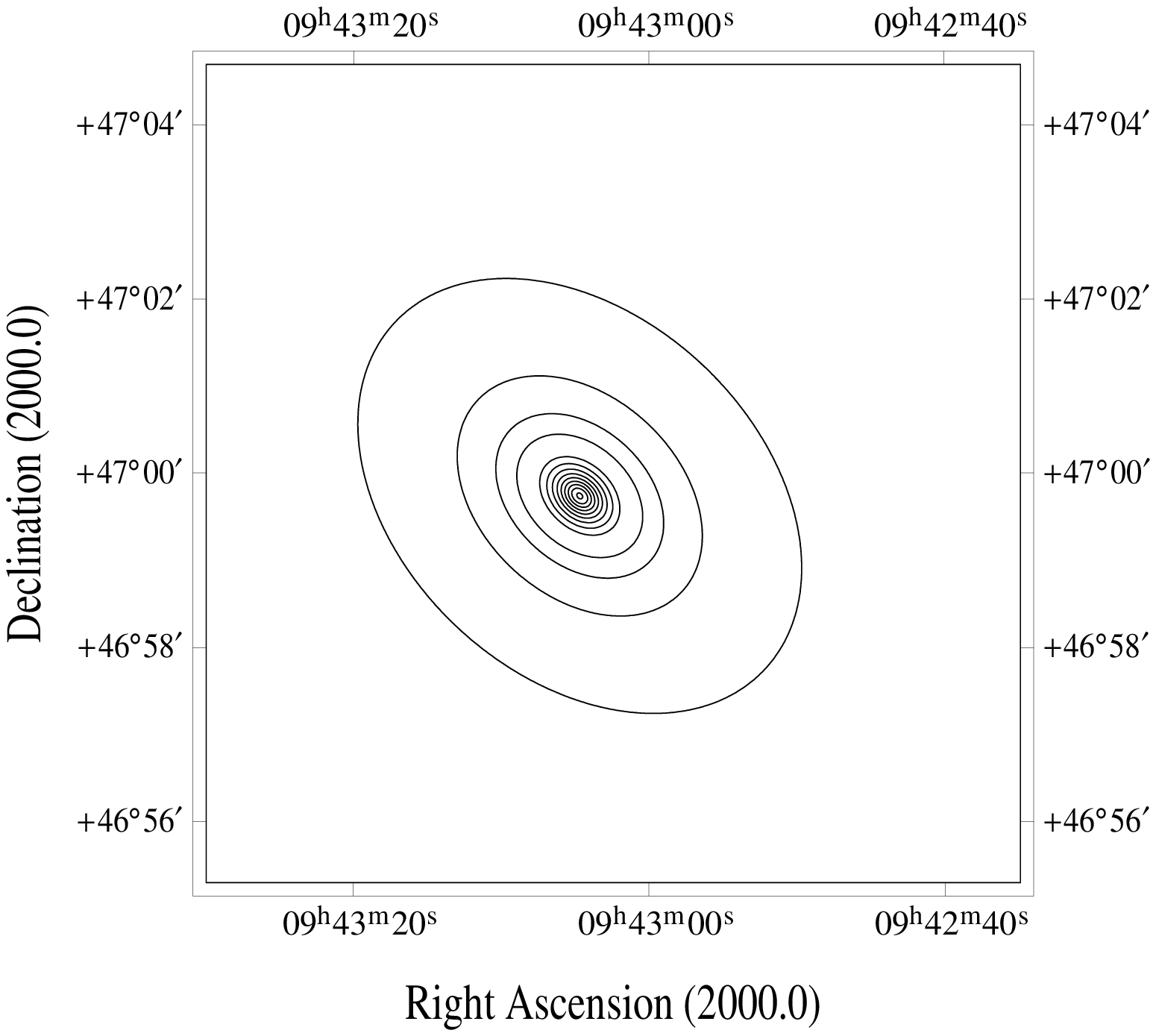,width=8.8cm,clip=} \\ 
\psfig{figure=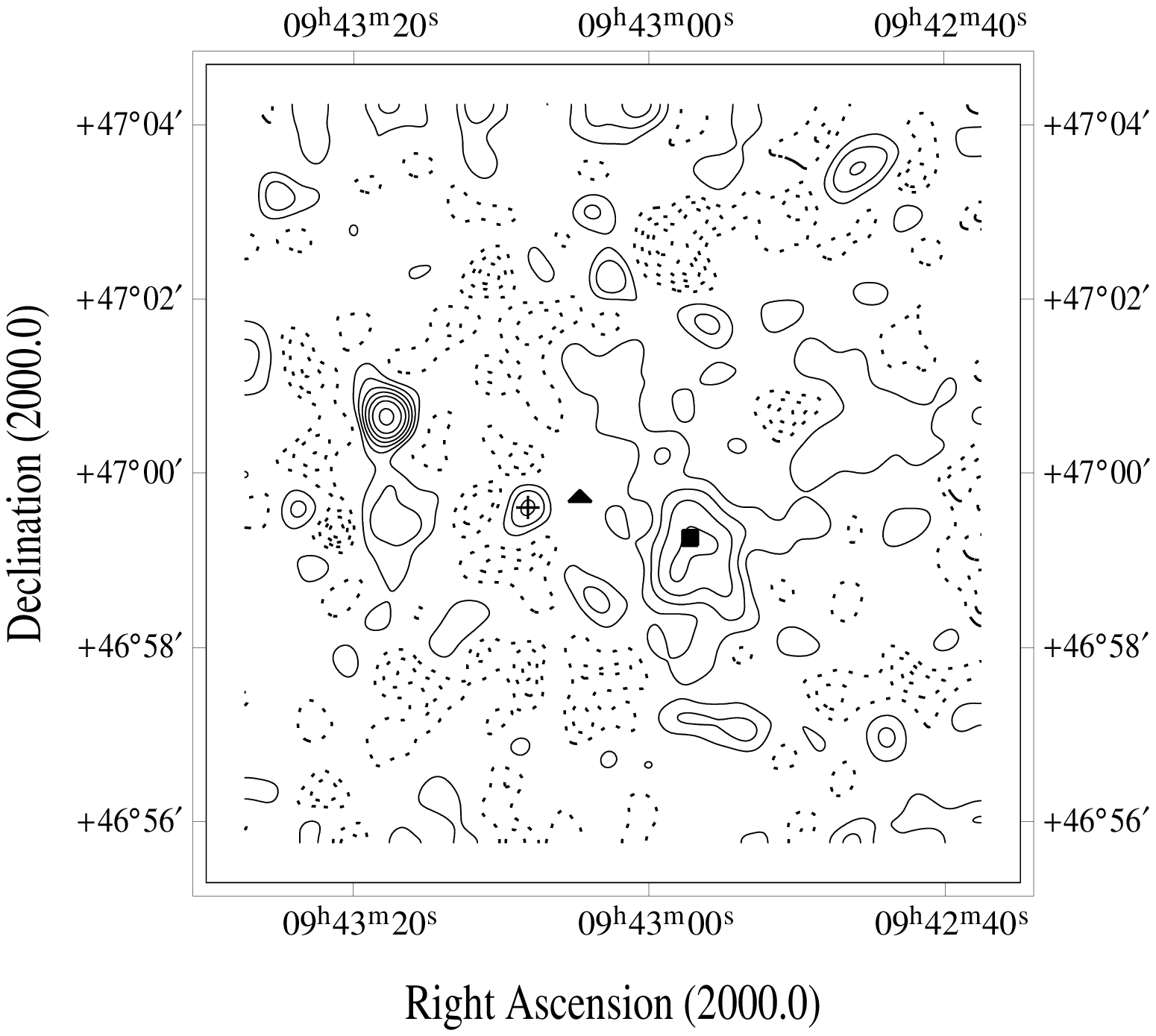,width=8.8cm,clip=} \\
\end{tabular}
\caption[]{Top: elliptical $\beta$ model (see Table 1); the contours are
logarithmically spaced with a spacing of 0.03.
The central
surface brightness is $1.3\times 10^{-2}$ cts/arcmin$^2$/sec. 
Bottom: Residual image after subtracting the elliptical $\beta$
model. The image is smoothed with a Gaussian filter of $\sigma=$10
arcsec. The contours represent the significance of the emission above
the $\beta$ model in units of $\sigma$. Dotted contours represent
negative values. The positions of M1 (square), M2 (triangle) and the
quasar at redshift two (plus sign) are indicated.
The substructure in the west around M1
has a significance of more than 4$\sigma$. It is clearly extended and
structured. The emission of the  z=2.055
quasar has a significance of more than 3$\sigma$.
}
\end{figure}

\subsection{Spectral analysis} \label{subsec-spec}

In order to complement the morphological analysis of CL 0939+4713 we 
perform a spectral analysis as well. 
For that we use data obtained by both ASCA/GISs (G2 and G3) and ASCA/SISs (S0
and S1) detectors and 
the ROSAT\-/PSPC. 

For composing the ASCA spectra the photons are extracted from a circle
centred on $9^h42^m55.5^s$, $46^{\circ}59'47''$ with a radius of 3.5
arcminutes and 4.5 arcminutes for the SIS and GIS,
respectively. It is necessary to extract the data from different
radii because of the different point spread functions of these
instruments. 

The GIS spectrum covers an energy range of 0.7-10 keV and is rebinned
to contain at least 50 photons per bin. The background is taken from
blank sky fields where the contaminating sources are masked (Ikebe 1995). 

The SIS data are difficult to analyse as they were taken with a
non-standard energy discrimination threshold. While normally
a lower energy discrimination threshold of 0.41 keV is used, these SIS data
were taken with a threshold of 0.55 keV. The correction of the
detector response seems to be inaccurate for this non-standard
threshold, e.g. if we use the SIS energy range down to 0.7 keV, we
find very high $n_H$ values not consistent with the PSPC results.
Therefore we have to restrict the SIS energy range for the spectral
fits. In Table 2 we list the fit results for two different SIS energy
ranges 3-6 keV and 1-10 keV. Obviously, these are both valid energy
ranges because they give very similar results.
The SIS data are rebinned to contain at
least 20 photons per bin and the standard Godard
background (see http://asca.gsfc.nasa.gov/docs/asca/abc\-/abc.html) is used.
In the ASCA images a possible contribution of
RXJ0943.0+4701 is not resolved, therefore this
emission can affect our results. But as RXJ0943.0+4701 has a softer
spectrum than  the cluster (see Sect. 2.5), its influence is expected 
to be very small on the ASCA spectra. 
Extrapolating the flux of RXJ0943.0+4701 as measured in the ROSAT/PSPC
observation to an energy band of 1-10 keV -- assuming a power law 
spectrum with a mean photon index of 2.5 (see Sect. 2.5) -- we find 
that the flux $F_{{\rm RXJ}0943}$(1-10keV) is less than 
1\% of the cluster flux.

The source spectrum of the ROSAT/PSPC is extracted from 
a circle centred between M1 and M2 ($9^h43^m01.6^s$, $46^{\circ}59'40''$)
with a radius of 2.3 arcminutes (the point spread function is even
smaller than the one of the ASCA/SIS).
We exclude the emission of RXJ0943.0+4701 ($9^h42^m56.8^s$,
$47^{\circ}00'47''$ with a radius of 50 arcseconds) 
since it is most likely unrelated to the cluster.
The background is taken from different parts of the same pointing. The
spectrum is rebinned to contain at 
least 16 and 25 photons, respectively. Fits with different background
realizations and different binning give very similar results.

We fit the spectra of all instruments simultaneously by folding for each
spectrum the
Raymond-Smith model with the response for the corresponding instrument
(see Fig. 4).
This procedure has the advantage of enabling us to constrain
different fit parameters at the same time. While the PSPC data can
determine the temperature only with very large uncertainty 
and cannot constrain the metallicity at all, they can
determine the hydrogen column density quite reliably, 
because the $n_H$ affects mainly the soft part of the spectrum. The GIS
data are ideal for determining the temperature, because the spectrum
extends to the highest energies (see Fig. 4). 
The SIS data are the ones that give good
constraints on the metallicity from the Fe K emission line, 
as the energy resolution of the SIS
is the best of all used detectors. Because all the fit parameters are
connected, it is the best to fit the spectra of all detectors 
simultaneously.    

The results of the spectral analysis are listed in Table 2. 
We show different simultaneous fit results, for two SIS energy ranges,
each one with the redshift
as free fit parameter as well as the redshift fixed to the optical value
of $z=0.41$. The results of all fits agree
well within the errors. The 
hydrogen column density is consistent with the Galactic value of
1.27$\times10^{20}$ cm$^{-2}$  
(Dickey \& Lockman 1990) within the uncertainty. 
In the following we use the mean values from all four simultaneous fits
and corresponding 
90\% confidence level
errors:
$T=7.6_{-1.6}^{+2.8}$ keV, $m=0.22_{-0.22}^{+0.24}m_{\odot}$ and
$n_H=0.9_{-0.4}^{+0.5}\times10^{20}$cm$^{-2}$. 

We find small discrepancies in the fluxes of the different
detectors: the flux determined by GIS  is about 15\% higher 
with respect to the flux determined by
SIS and about 25\% higher with respect to the PSPC flux,
which can be explained only partially by the different
extraction radii and variability of the source RXJ0943.0+4701 (see
Sect. \ref{subsec-m3}). 

%
%
%
%
\begin{table*}[htbp]
\begin{center}
\begin{tabular}{|c|c|c|c|c|c|c|c|}    
\hline
& & & & & & & \\
Instruments&SIS energy range in keV&$ k_{\rm B} T$ in keV& $m/m_{\odot}$ & $n_H$ in $10^{20}$cm$^{-2}$& $z$&d.o.f. &$\chi^2/$d.o.f.\\
& & & & & & &\\
\hline
& & & & & & &\\
GIS/SIS/PSPC & $3-6$  &  $7.4^{+2.4}_{-1.4}$ & $0.23^{+0.23}_{-0.20}$ 
	& $0.9_{-0.4}^{+0.5}$&$0.38^{+0.07}_{-0.04}$&81&0.97\\
GIS/SIS/PSPC & $3-6$  &  $8.1^{+2.3}_{-1.7}$ & $0.20^{+0.21}_{-0.20}$ 
	& $0.8_{-0.3}^{+0.5}$&$0.41$(fixed)&82& 0.97\\
GIS/SIS/PSPC & $1-10$  &  $7.2^{+1.6}_{-1.1}$ & $0.25^{+0.21}_{-0.19}$ 
	& $0.9_{-0.4}^{+0.5}$&$0.38^{+0.05}_{-0.03}$&139&0.92\\
GIS/SIS/PSPC & $1-10$  &  $7.6^{+1.6}_{-1.2}$ & $0.19^{+0.20}_{-0.16}$ 
	& $0.9_{-0.4}^{+0.5}$&$0.41$(fixed)&140& 0.92\\
PSPC only    &        &$5.7^{+\infty}_{-2.7}$ & 0.23(fixed) &
          $1.0_{-0.5}^{+0.6}$& 0.41(fixed)& 18 & 0.86\\ 
& & & & & & & \\
\hline
\end{tabular}
\end{center}
\caption{Results of the simultaneous spectral fits of the data from
both ASCA/SIS detectors, from 
both ASCA/GIS detectors and from the ROSAT/PSPC. A fit of the
ROSAT/PSPC data alone is shown as well. The columns show the
temperature $k_{\rm B} T$ in keV, 
the metallicity $m$ in solar units, the 
hydrogen column density $n_H$,
the X-ray determined redshift $z$, the degrees of freedom of the fit
and the reduced $\chi^2$. 
The errors are 90\% confidence level. 
For the uncertainty in redshift a systematic error of 1\% in the  
ASCA energy calibration is included.    
}
\end{table*}

\begin{figure} \label{fig-spec}
\begin{tabular}{c}   
\psfig{figure=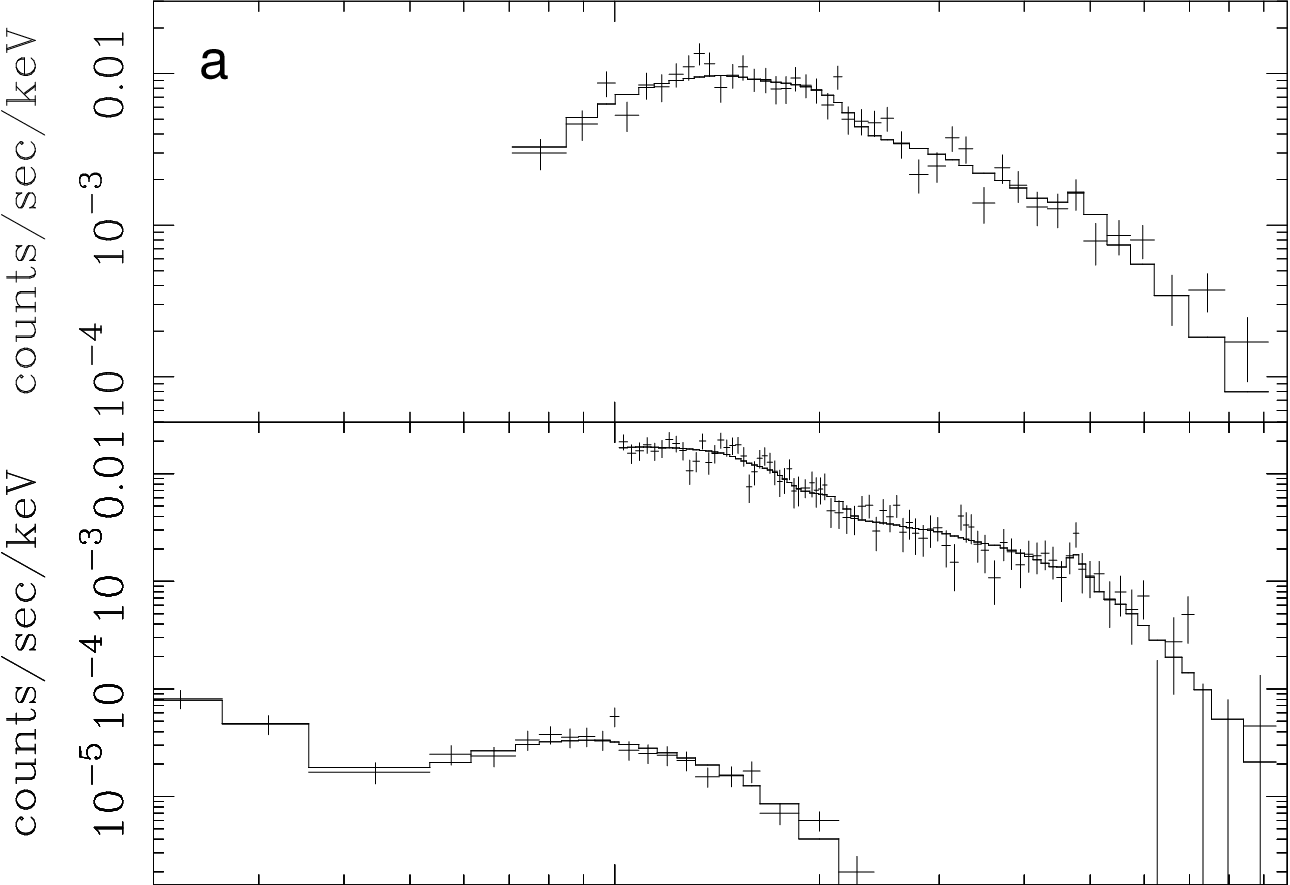,width=8.8cm,angle=-90.,clip=} \\ 
\psfig{figure=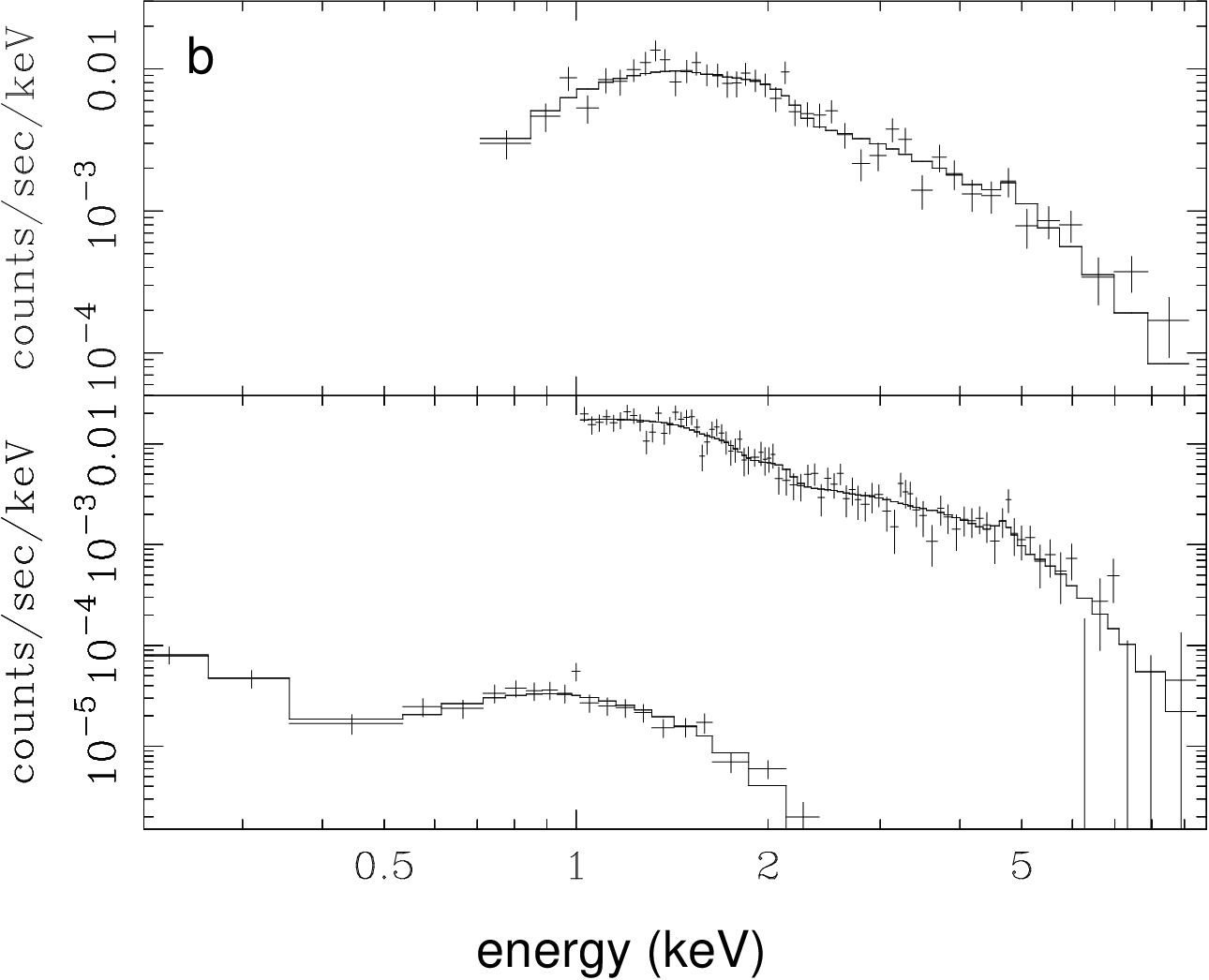,width=8.8cm,angle=-90.,clip=} \\ 
\end{tabular}
\caption[]{X-ray spectra of CL 0939+4713 obtained with the ASCA/GIS,
ASCA/SIS as well as the ROSAT/PSPC with the redshift as a free
fit parameter (a) and with fixed redshift (b). Each panel shows the GIS
spectrum (top), the SIS spectrum for the energy range of 1-10 keV 
(middle) and the PSPC spectrum
(bottom). The normalisation of the PSPC spectrum is shifted
arbitrarily for the image. 
A Raymond-Smith model 
(folded with the response of the corresponding instrument)
which gives the best fit for the GIS, SIS and PSPC 
spectra simultaneously is superposed.
The contamination due to RXJ0943.0+4701 is subtracted from the PSPC spectrum. 
}
\end{figure}

For comparison with the temperature found previously in a ROSAT/PSPC analysis
by Schindler \& Wambsganss (1996) we show in Table 2 also the fit of
the PSPC data only. The temperature of T$=5.7^{+\infty}_{-2.7}$ keV 
is somewhat lower than in the
combined fits, but the upper limit is unconstrained.
The temperature of Schindler \& Wambsganss (1996) $T=2.9^{+1.3}_{-0.8}$
keV obtained without excluding  RXJ0943.0+4701 was much lower.
The explanation is that the
ROSAT/PSPC spectrum was contaminated by the soft contribution of the 
variable source RXJ0943.0\-+4701 and therefore the best fit yielded a
lower temperature.

\subsection{Luminosity} \label{subsec-lum}

The X-ray emission of the galaxy
cluster can be traced out to 2.5 arcminutes
(almost 1 Mpc) in the HRI observation. 
Within this radius we find 1060 source counts after 
background subtraction (the point source P1 lies outside this circle; 
the emission of the quasar is  excluded.)
This corresponds to a count rate of 0.023 HRI counts/s. 

Using the temperature $T=7.6_{-1.6}^{+2.8}$ keV
and the hydrogen column density
$n_H=0.9_{-0.4}^{+0.5}\times10^{20}$cm$^{-2}$ 
of the spectral fits
we derive a luminosity in the ROSAT band (0.1-2.4
keV) of $6.4_{-0.3}^{+0.7}\times 10^{44}$erg/s,
corresponding to a bolometric luminosity of
$1.6_{-0.3}^{+0.5}\times 10^{45}$erg/s. 
%

%
\subsection{Gas mass, total mass and iron mass} \label{subsec-mass}

A mass determination from the X-ray data
for this cluster is very difficult, because one
has to assume spherical symmetry, which is certainly a bad approximation
for CL 0939+4713. Just for giving a rough estimate of the mass we restrict
the mass analysis to the central parts of the two subclusters assuming
that each subcluster is spherically symmetric. The
outer radius of 41 arcseconds (265 kpc)
is chosen as half of the distance between the
subclusters. For the analysis we use the spherically symmetric
$\beta$-models described in Sect. \ref{subsec-morph} 
and a constant temperature of 7.6 keV for both subclusters. 

Figure 5 shows the integrated gas and total masses. 
At a radius of 265 kpc the integrated gas masses in the subclusters M1
and M2 are very similar:
$M_{\rm M1,gas}(r\le 265$kpc$) = 4.1\times 10^{12} \msol$ and
$M_{\rm M2,gas}(r\le 265$kpc$) = 4.2\times 10^{12} \msol$, respectively. 
Assuming a constant temperature  and hydrostatic equilibrium 
we can also determine the total masses
of the subclusters: 
$M_{\rm M1,tot}(r\le 265$kpc$) = 6.0_{-1.6}^{+2.3}\times 10^{13} \msol$
and
$M_{\rm M2,tot}(r\le 265$kpc$) = 7.4_{-2.0}^{+2.9}\times 10^{13} \msol$,
respectively. The errors come mainly 
from the uncertainty in the temperature (see
Sect. \ref{subsec-spec}), which are $_{-22\%}^{+35\%}$. 
An additional uncertainty is introduced by the internal
substructure of the subclusters
which tends to cause an underestimation of the mass (Schindler 1996).
The errors on the mass 
include another uncertainty of 15\% error coming from this effect, 
from possible temperature
gradients and from possible temperature differences between the subclusters.

\begin{figure}
\psfig{figure=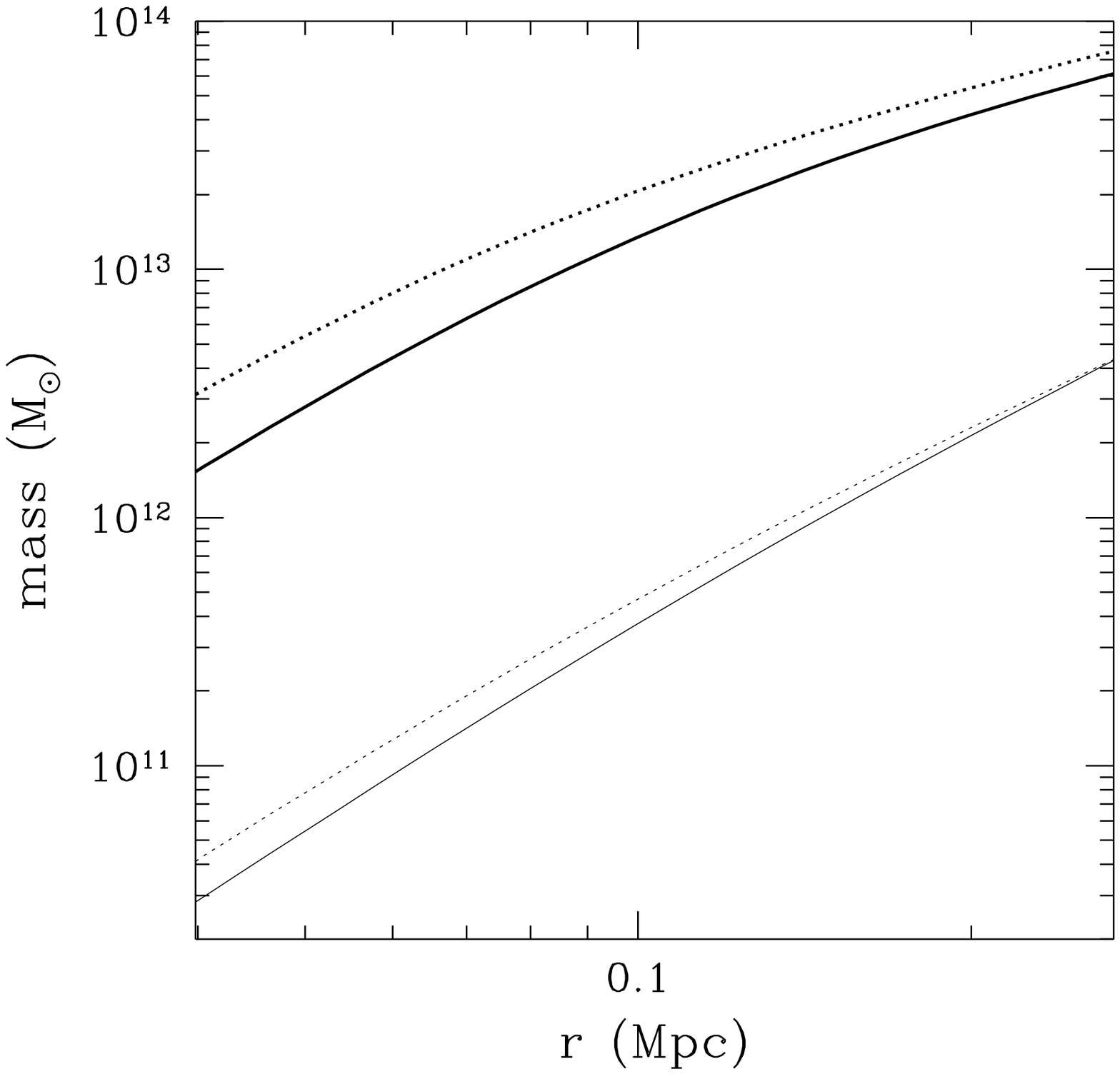,width=8.8cm,clip=} \label{fig-mass}
\caption[]{Profiles of the integrated gas mass
(thin lines, bottom)
and the total mass (thick lines, top) of the subclusters M1 (solid lines)
and M2 (dotted lines). 
}
\end{figure}

With these numbers we 
find gas mass fractions of $(7\pm2) \%$ and $(6\pm2)\%$ 
for M1 and M2,
respectively. These fractions are very small compared to other
clusters where one finds usually gas mass fractions between 10\% and
30\% (B\"ohringer 1995).  

For a comparison with the total mass estimate from gravitational lensing we
determine the mass in the region were the HST field and our two
circles around the subcluster centres overlap. The HST field is much
smaller than the HRI field so that only part of the two circles
are covered by the
HST field. As M1 and M2 are close to the border of the HST field we
use only the sectors of the circles which have overlap with the HST field:
angles $+128^\circ - +294^\circ$ (N over E) centred on M1 and angles
$-90^\circ - +90^\circ$ centred on M2. The total overlapping area
is only about 1.4 arcmin$^2$. For the comparison we have to determine
the surface mass density i.e. projecting
the mass onto a plane. For doing this we have to extrapolate 
and project the 
$\beta$ model. As the outer integration radius in the line of sight 
we choose
1 Mpc, which is a compromise between extrapolating too far into non-relaxed
regions and summing 
most of the mass. Integrating out to a radius of 1.5 Mpc would increase the
mass only be 3.5\% which is negligible considering the error.
With this procedure we find a projected mass of $1.0^{+0.4}_{-0.3}\times
10^{14}\msol$ in the overlapping area. The error is again the same
percentage as used for $M_{\rm M1,tot}$ and $M_{\rm M2,tot}$.

The mass from gravitational lensing as calculated by Seitz et
al. (1996) in the same overlapping area is $2.3\times10^{14}\msol$. 
For this number a mean redshift of the background
galaxies of $z=1.0$ is assumed. If one would use a different redshift
$z=0.8$ and $z=1.5$, the mass would be $2.8\times10^{14}\msol$ and
$1.8\times10^{14}\msol$, respectively. As this redshift is the largest
uncertainty in the lensing mass estimate, it remains a 
discrepancy of a factor 2 -- 3 between the X-ray and the lensing mass.

For determining the iron mass $M_{Fe}$ in the intra-cluster gas
we use the metallicity 
from the spectral
fits $m=0.22_{-0.22}^{+0.24}m_{\odot}$. With this we calculate the ratio 
of $M_{Fe}/M_{gas} = 3.2^{+3.4}_{-3.2}\times10^{-4}$. For the
determination of the gas mass we use the two spheres of 265 kpc radius
around the two subclusters. The iron mass in this volume 
is $M_{Fe}=2.7\times10^9\msol$. Within the same radii in the optical field
are 26 galaxies. 
Estimating the total blue luminosity including an
extrapolation to faint galaxies we find about $L_B = 6\times10^{11}\lsol$.
This results in an iron mass to light ratio $M_{Fe}/L_B \approx
5\times10^{-3}\msol/\lsol$. 

For comparison we do the same also in a
larger sphere of 584 kpc around M1 (the maximum radius 
completely covered by
the optical field) by extrapolating the $\beta$-profile.
The subcluster around M2 is neglected in this case.
For this case we find $M_{Fe}=7.7\times10^9\msol$, 75 galaxies with $L_B =
15\times10^{11}\lsol$, and again $M_{Fe}/L_B \approx 
5\times10^{-3}\msol/\lsol$.
As the two numbers for the iron mass to light ratio 
are so similar, we can be sure that we do not make an error
by using a restricted volume and by not taking into account the gas in
the line-of-sight outside the circles.

The iron mass to light ratio $M_{Fe}/L_B \approx
5\times10^{-3}\msol/\lsol$ is
smaller than the ones typical for
rich clusters of 0.01-0.02 (Renzini et al. 1993), although one has to
keep in mind the errors only coming from the metallicity 
are already $\pm100\%$.

%
\subsection{The variable X-ray source RXJ0943.0+4701} \label{subsec-m3}

A comparison of the PSPC and the HRI observations shows that the
source
RXJ0943.0+4701 is  not present in the HRI image (Fig. 6).
For a determination of the flux of RXJ0943.0+4701 in the PSPC observation 
we simulate the PSPC point spread function, 
smooth it with a Gaussian of $\sigma=20$arcsec and subtract it
from a PSPC image with the same smoothing. 
We vary the number of counts in the point spread function model 
until  the level of the residuals is comparable to the 
diffuse cluster emission at this position.

With this method we find that RXJ0943.0+4701 has about 50 counts (broad band) 
above the cluster emission in the PSPC observation. 
This corresponds to about $3.5\times10^{-3}$ cts/s for the PSPC 
detector. 
With the above mentioned value of the Galactic $n_H$ and 
a power law with photon index 2.3 this count rate corresponds to 
a flux in the ROSAT band (0.1-2.4 keV) of
$f_{\rm ROSAT, Nov.91}=4\times 10^{-14}$ erg/cm$^2$/s. 

In the HRI image appears very faint emission 
near the position of RXJ0943.0+4701  with very low significance 
(only 1.3$\sigma$) which could as well be a statistical fluctuation. 
If there is any real emission at this position it cannot contain
more than 5 counts above the cluster plus background emission;
this corresponds to a source countrate $\le 1\times10^{-4}$ HRI cts/s. 
With the same
assumptions as above this count rate corresponds to a flux in the ROSAT
band $f_{\rm ROSAT, Oct.96}\le 5\times 10^{-15}$ erg/cm$^2$/s. 
Given the smaller point spread function of the HRI which would give
the (point) source a higher significance above the background 
and the diffuse cluster emission we conclude that the decrease in 
luminosity of RXJ0943.0+4701 in the ROSAT band is at least a factor of
10. This decrease  can be either due to a general decrease in
luminosity or due to a shift
of the emission out of the ROSAT energy range.

The non-detection of RXJ0943.0+4701 in the HRI observation cannot be
caused by the
fact that the HRI is slightly less sensitive in the soft band compared
with the PSPC (ROSAT User's Handbook
http://www.xray.mpe.\-mpg.de/rosat/doc/ruh), because
the hardness ratio at the position of RXJ0943.0+4701
(hard-soft)/(hard+soft)\-$=-0.08$, i.e. there are about the same number of
counts in the soft and in the hard band.

\begin{figure*}
\begin{tabular}{ll}   
\label{fig-comb}
\psfig{figure=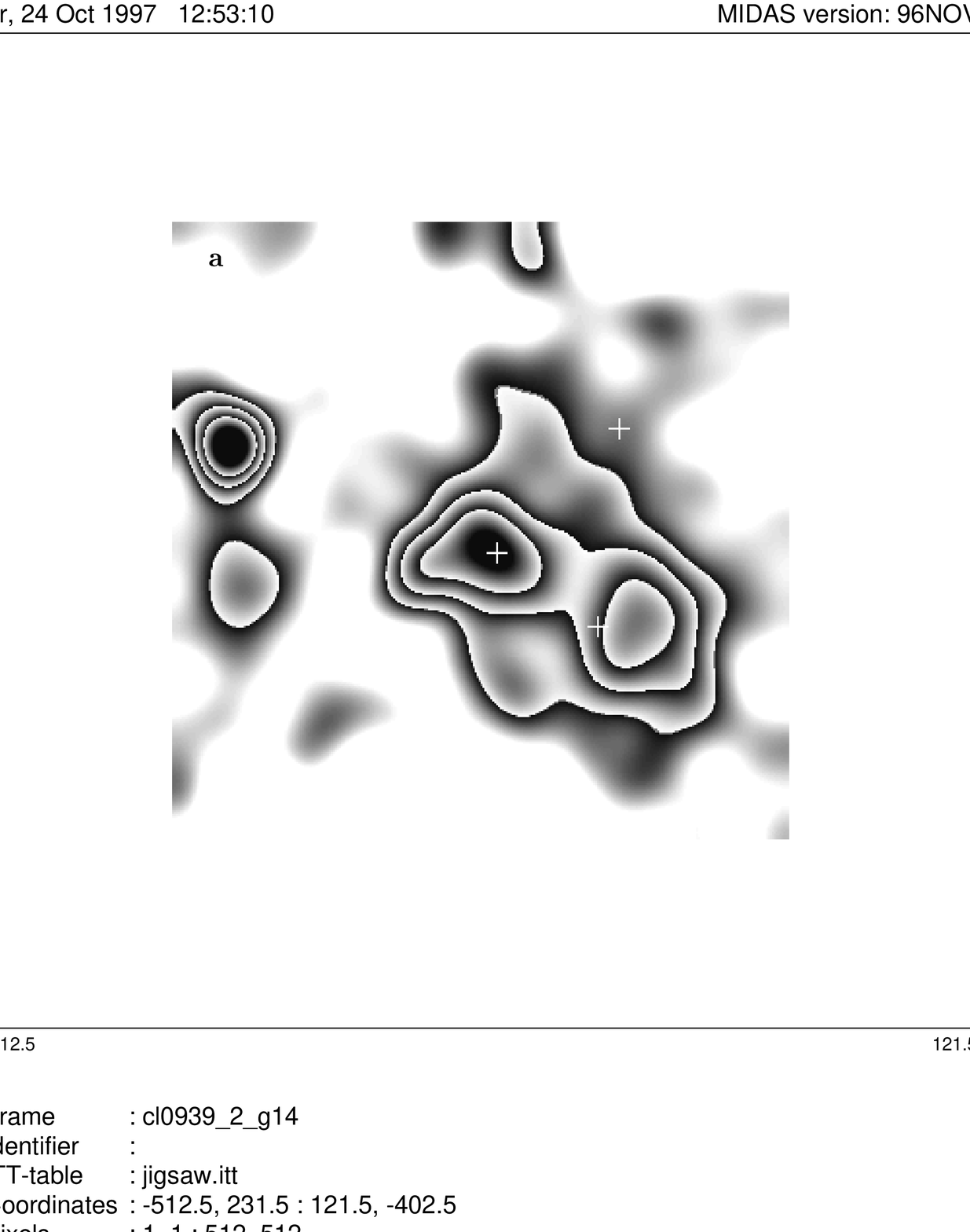,width=7cm,clip=} &
\psfig{figure=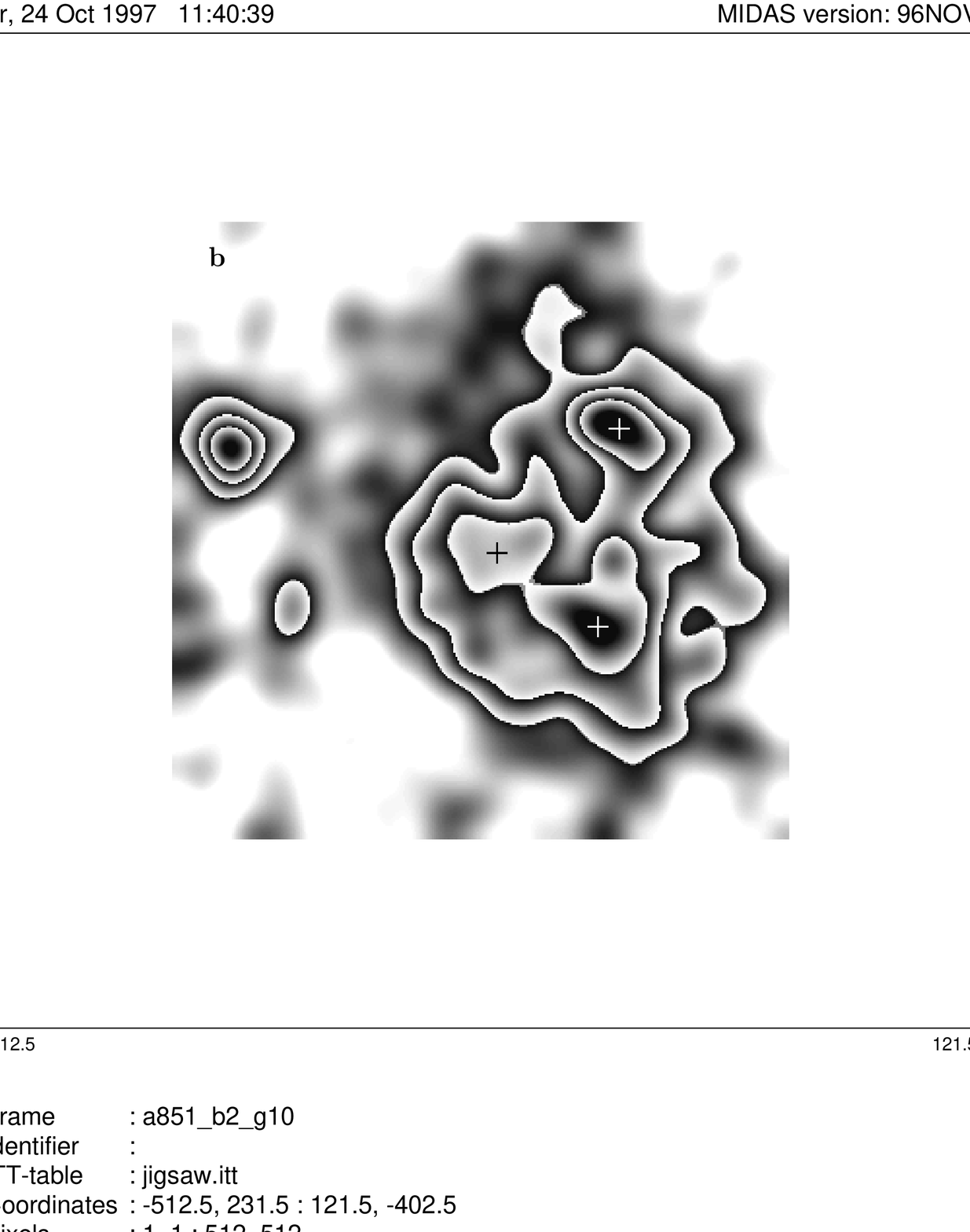,width=7cm,clip=} \\
\psfig{figure=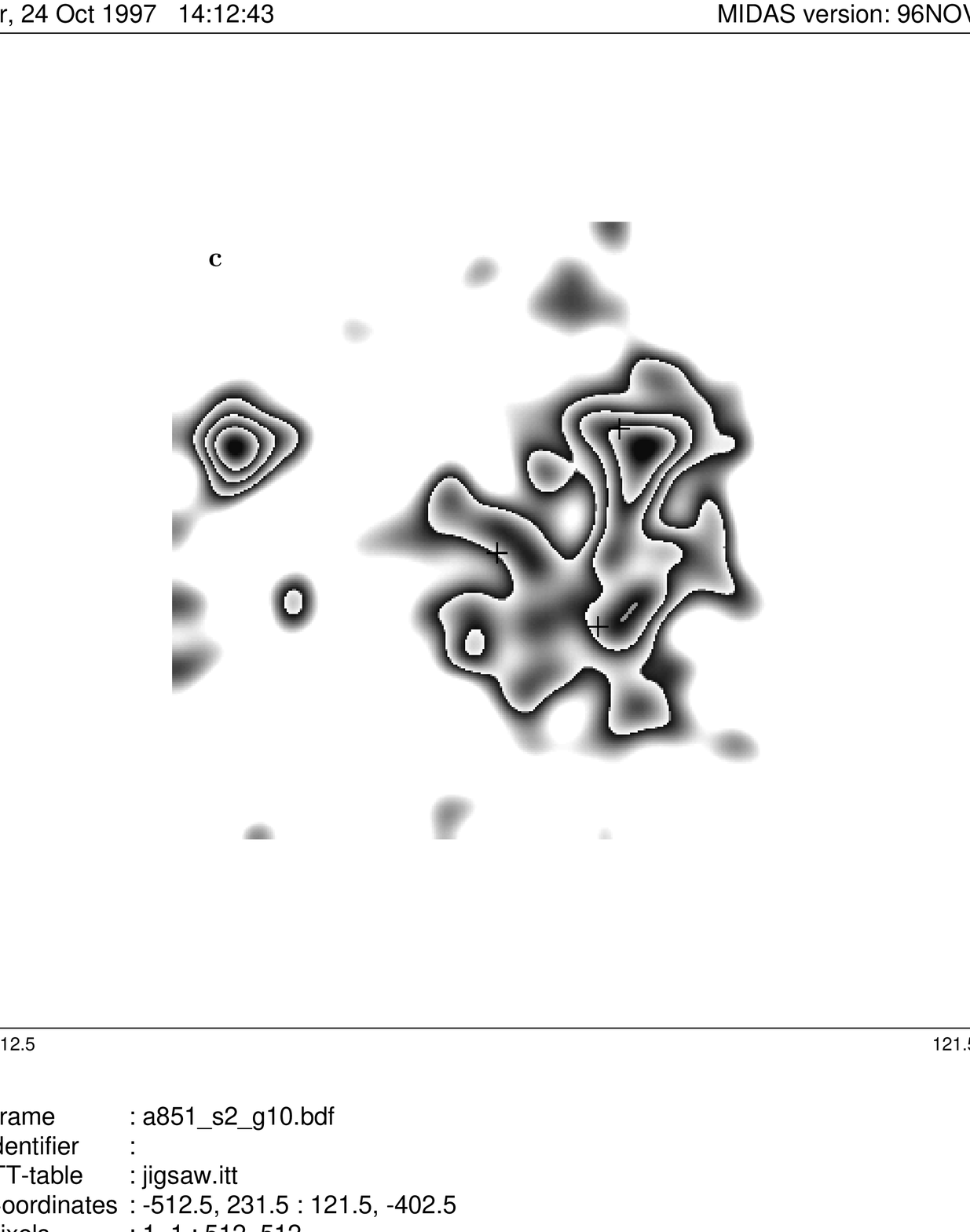,width=7cm,clip=} &
\psfig{figure=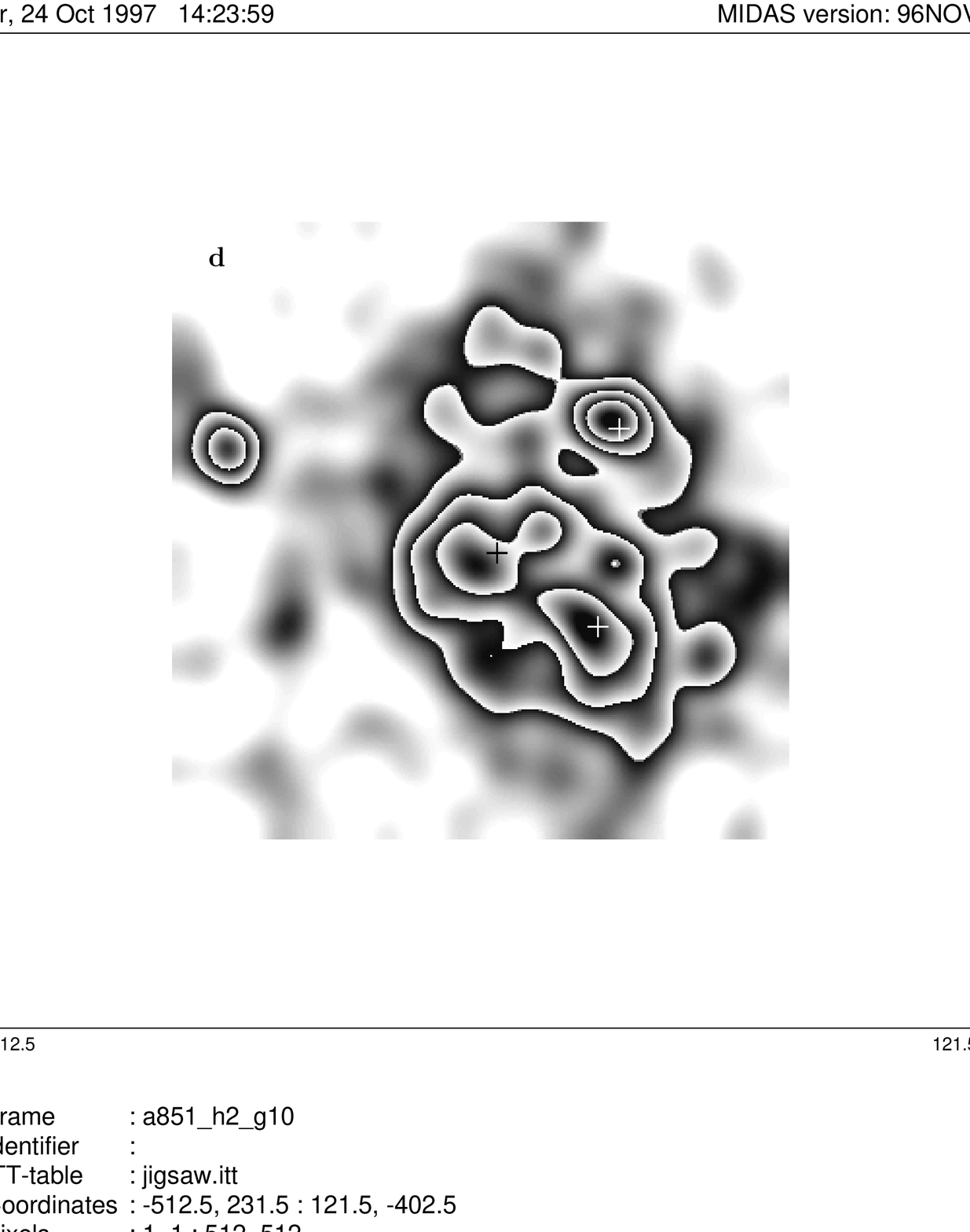,width=7cm,clip=} \\
\psfig{figure=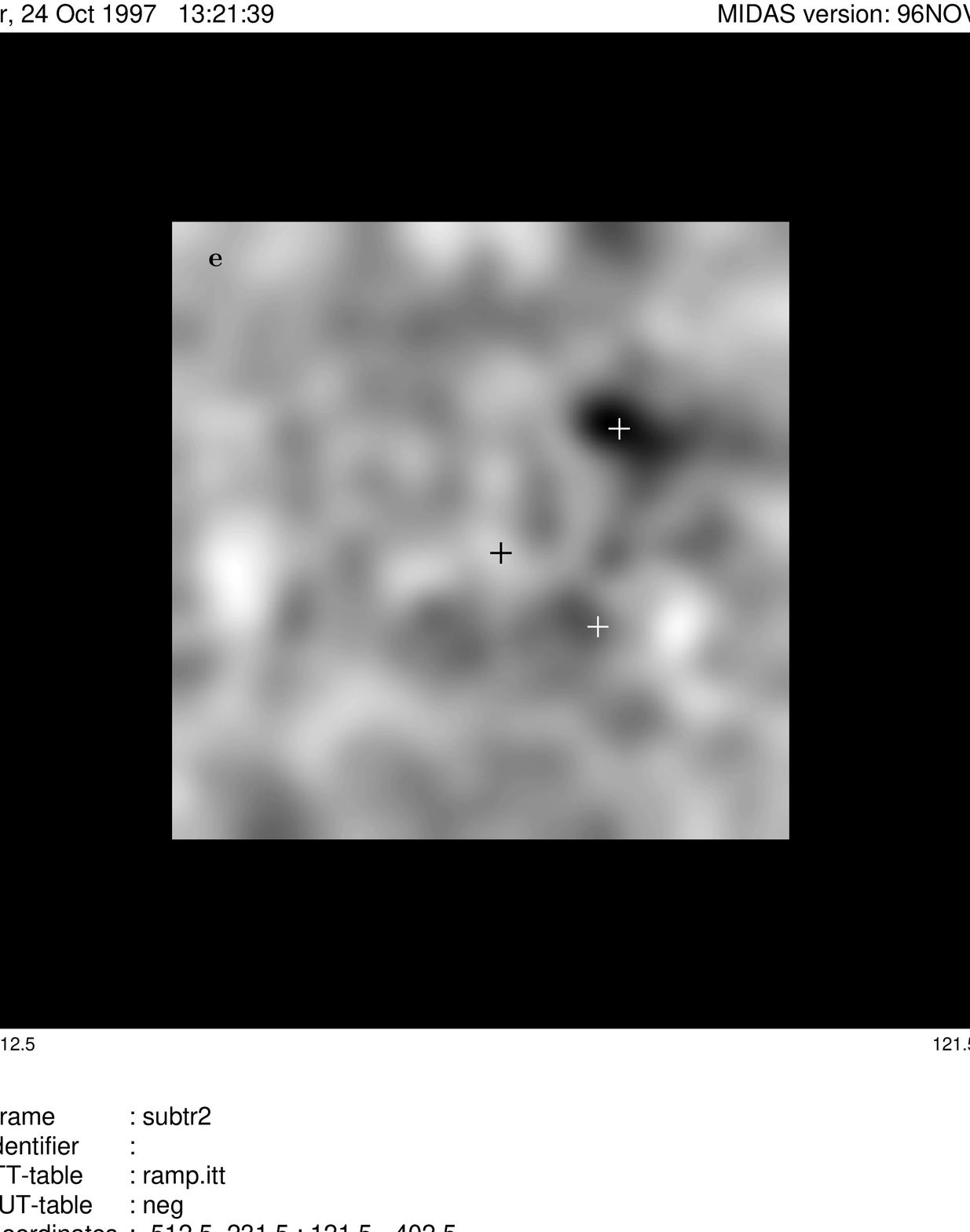,width=7cm,clip=} & \\
\end{tabular}
\caption[]{
Morphology of CL 0939+4713 as seen in an HRI image (a) and in a PSPC
image (b) taken 5 years earlier. 
The HRI image is smoothed
with a Gaussian of $\sigma=14$arcsec so that the resolution
is about equivalent to the PSPC observation.
b) shows the PSPC image in broad
band which is the corresponding band to the HRI image.
The PSPC images are smoothed with a Gaussian of $\sigma=10$arcsec.
In each of the panels the positions of the maxima of the PSPC 
broad band image are marked with crosses. The comparison 
shows that the northern maximum of the PSPC observation 
(RXJ0943.0+4701) has disappeared in the HRI observation. 
The relative brightness of the maxima M1 
and M2 seems to have changed, but this difference
is within the expected statistical
fluctuations. 
c) PSPC soft (0.1-0.5 keV) and d) PSPC hard
(0.5-2.0 keV)
band images are shown to demonstrate that RXJ0943.0+4701 is 
softer than the rest of the cluster emission. The positions of the
northern X-ray maximum are not exactly coincident in the soft and the
hard band.
e) Subtracted image:  HRI (a) -- PSPC (b) image. The most obvious
feature is the black 
minimum at the position of RXJ0943.0+4701. The minimum
is caused by 
the missing RXJ0943.0+4701 emission in the HRI image. Also the small
intensity variations of 
M1 and M2 are visible (M1 slightly fainter, M2 slightly brighter in
the HRI image). The emission of the quasar at redshift two is also
visible as a faint white region left of M2, but this excess
emission in the HRI image is not significant. 
The images all have a size of 5 arcmin at a side.
}
\end{figure*}

In the ROSAT All-Sky Survey (Voges et
al. 1996) CL 0939+4713 was observed in November 1990 for two days.
 Unfortunately,
the exposure time is too low to see any morphological details.  
The Survey countrate of the whole cluster region is within the
errors in agreement with both the PSPC and the (converted) HRI count rate.
A lightcurve of the Survey observation shows no indication of
variability (Boller \& Voges, private communication).

Unfortunately, RXJ0943.0+4701 cannot be resolved with  ASCA, i.e. one
cannot distinguish the cluster emission from emission coming from
RXJ0943.0\-+4701. Therefore, we have no information about the flux of
RXJ0943.0\-+4701 at the time of the ASCA observation.

For examining the short-term variability of RXJ0943.0\-+4701 we try to 
derive a lightcurve from the PSPC observation. The source was
observed in eight intervals within 72 hours. The intervals have
exposure times between 1300s and 2400s. Figure 7 
shows this attempt of a lightcurve. 
It is consistent with a constant emission over 3 days. 

A spectral fit to the RXJ0943.0+4701 region alone is very difficult because 
of the small number of photons. 
In an attempt to get a least a rough idea we fit
the spectrum using two different kinds of background: 1) only detector
background and 2) detector background plus cluster emission. 
A fit with a Raymond \& Smith (1977) model yields temperatures 
of $T=1.1$keV and $T=0.3$keV, respectively, for the two different
background models. Although these results have very large errors they
show that the RXJ0943.0+4701 emission is considerably
softer than the 
cluster emission (compare also Fig. 6). 
A fit with a power law to
the same spectra yields photon indices of 2.2 and 2.9, respectively
(again with very large errors).

\subsection{Identification of RXJ0943.0+4701} \label{subsec-opt}
Since the source RXJ0943.0+4701 appears to be variable and hence very
interesting, 
we try to find the optical counterpart of the X-ray emission at
RXJ0943.0+4701. Unfortunately, the northern part of the cluster was
never observed with HST.
A ground based I-band image of the field around RXJ0943.0+4701 is shown in 
Fig. 8. 
The position of RXJ0943.0+4701 is not very well determined in the PSPC
observation  
for several reasons. First, the maxima in different bands are not
exactly coincident (see Fig. 6 and 8).
Second, there is no other point source near the cluster with an obvious
optical counterpart which
could be used to correct for a possible pointing offset of the ROSAT
telescope. Therefore, we use the HRI image for the correction. In this
image the quasar at z=2.055 is used to determine the pointing offset of
the HRI. The pointing offset is only 1 arcsec. After correcting for
this offset we use the point source P1 (see Fig.1)
present in both, HRI and PSPC image, to correct for the PSPC offset, which
is relatively large, 11 arcsec.

\begin{figure}
\psfig{figure=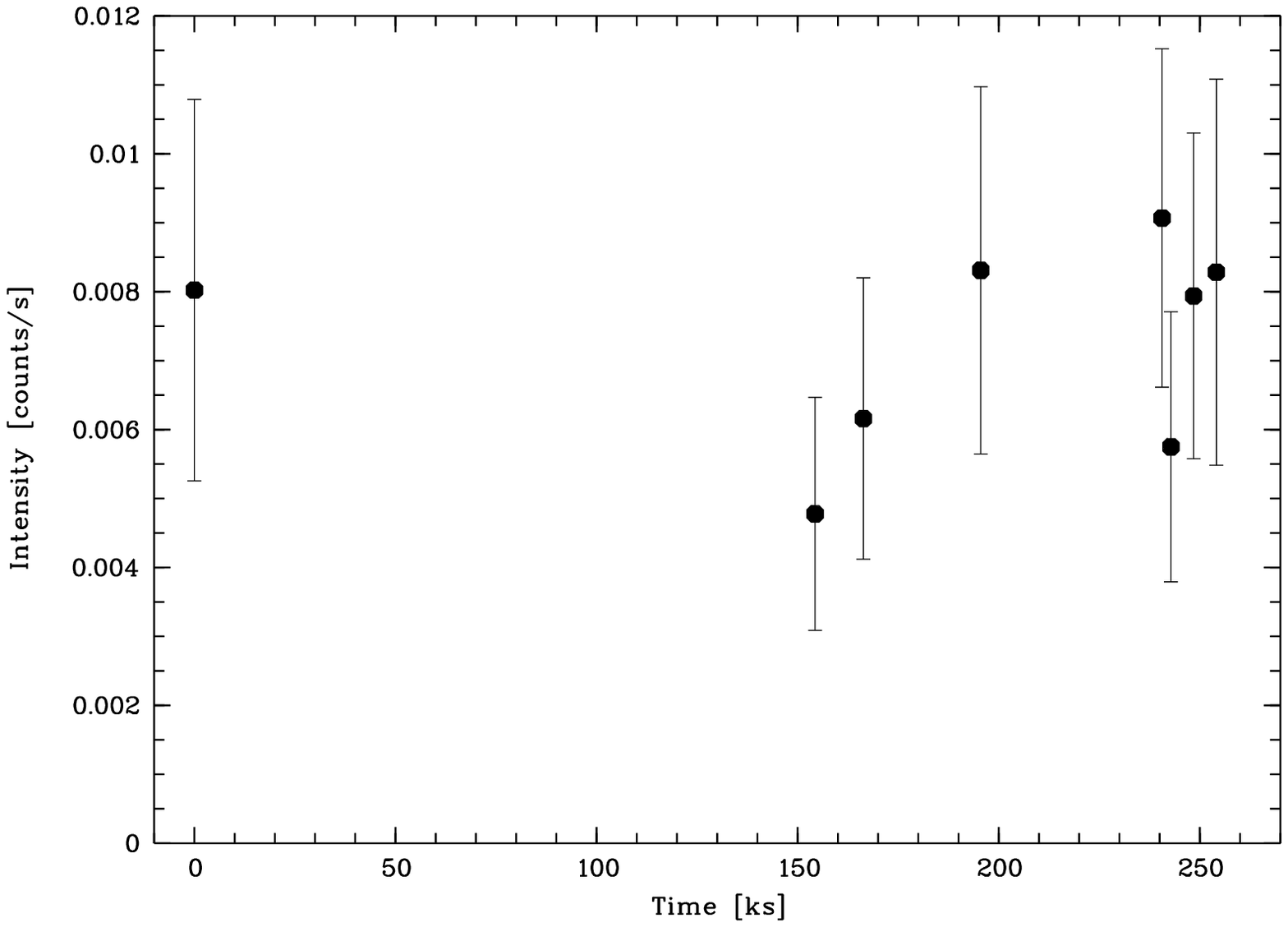,width=8.8cm,clip=}  \label{fig-light}
\caption[]{Lightcurve of RXJ0943.0+4701 from the ROSAT/PSPC observation. It is
consistent with a constant brightness over the 
observing interval of   
72 hours. The observation was carried out in eight intervals with
exposure times  between 1300 sec and 2400 sec. 
The lightcurve is background 
subtracted, the cluster emission at the position of RXJ0943.0+4701 is
not excluded. As the cluster emission should be constant it should give
only an offset to the curve. Because of the limited number of photons
the error bars are quite large. 

}
\end{figure}

The corrected positions of RXJ0943.0+4701 are marked in the optical
image (Fig. 8). In the region 
of these positions are 8 possible optical 
counterparts with R magnitudes brighter than 22$^m$. 
These objects are marked with numbers in Fig. 8.
We measure the brightness of all these objects using fixed aperture
photometry with an aperture of 3.5 arcsec on the deep cluster images
taken with the 12 filters (Belloni et al. 1995). 
The R magnitudes of all these counterpart candidates are listed
in Table 3 along with their classification from the SEDs plus
morphological analysis. 
Figure 9 shows the observed fluxes and the SEDs
that we assign by eye as the best fit. 
Although this is admittedly a quite qualitative classification, we can 
still gain valuable information from it.      
Using all this information we gather:
of the possible optical counterparts two are stars,
five are galaxies and one is a  blue compact object (\#130).

\begin{table*}[htbp]
\begin{center}
\begin{tabular}{|c|c|c|l|c|}    
\hline
& & & &\\
object & $\alpha$(2000)&$\delta$(2000)& R magnitude & probable identification \\
& & & &\\
	\hline
& & & &\\
\#94   & 09 42 59.1 & 47 01 04 & 20.37 & spiral \\
\#99   & 09 42 58.2 & 47 01 02 & 19.75 & elliptical/Sa at z=0.30-0.35 \\
\#115  & 09 42 55.8 & 47 00 56 & 19.26 & M0-M2 or K6-K8 star \\
\#118  & 09 42 59.3 & 47 00 56 & 21.05 & Sb, possibly cluster member \\
\#130  & 09 42 57.2 & 47 00 50 & 20.60 & blue compact object, slightly larger than PSF \\
\#134  & 09 42 59.0 & 47 00 48 & 21.63 & possibly elliptical  and cluster member \\
\#158  & 09 42 57.5 & 47 00 35 & 21.61 & Sb-Sc, possibly cluster member \\
\#166  & 09 42 58.1 & 47 00 29 & 18.48 & M0-M2 or K6-K8  star \\
& & & &\\
	\hline
\end{tabular}
\end{center}
\caption{List of possible optical counterparts of the X-ray source
RXJ0943.0+4701 and probable identifications.  Numbers are as in Fig.
8. 
}
\end{table*}

\begin{figure*}
\psfig{figure=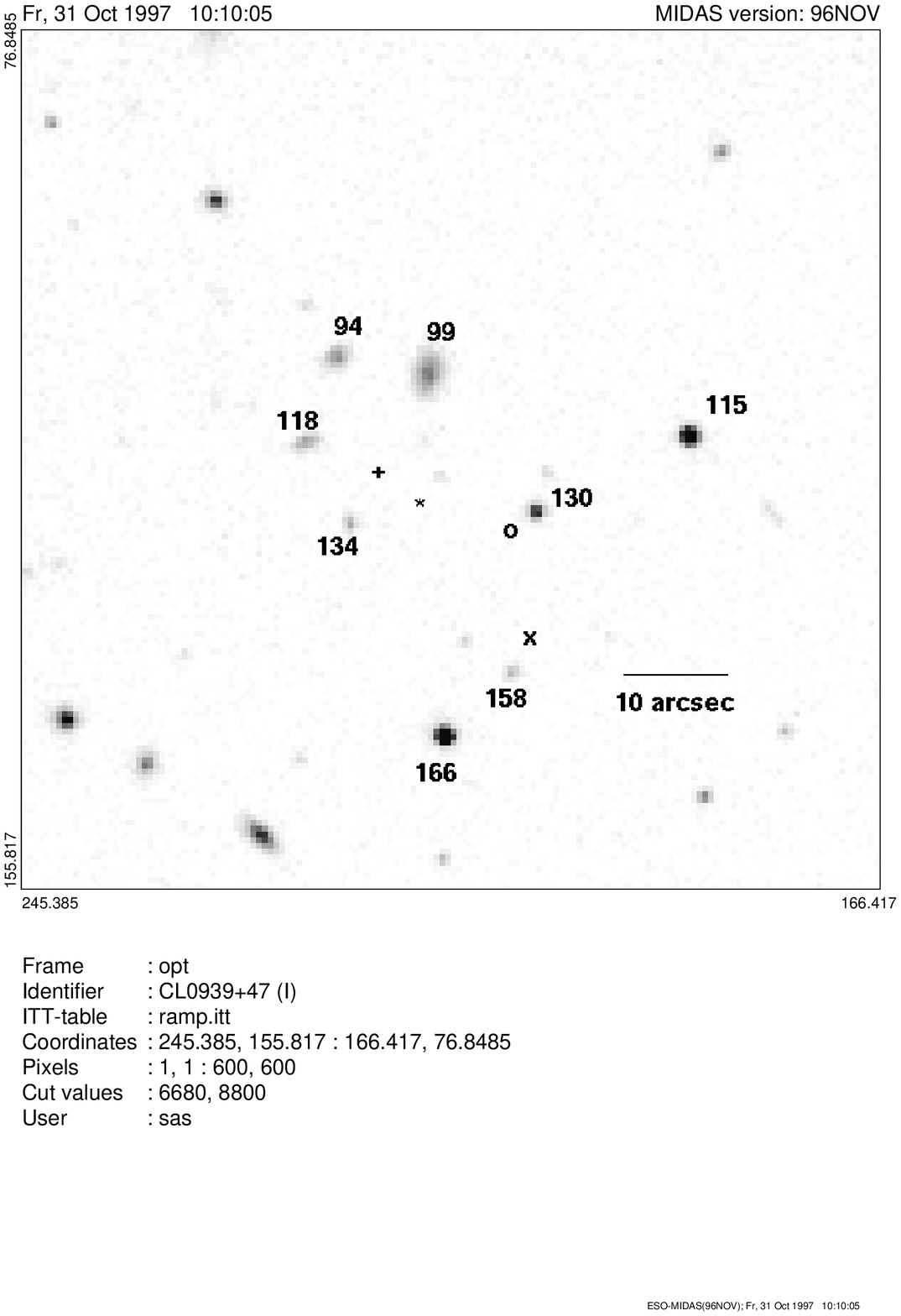,width=12.0cm,clip=}  \label{fig-opt}
\caption[]{Optical image in the region around RXJ0943.0+4701. 
The image has a size of about 80 arcsec by 80 arcsec
(east is left) and is taken in the I-band filter. 
The possible optical counterparts of RXJ0943.0+4701 are
marked with numbers (see Table 3). The position of the X-ray source 
RXJ0943.0+4701 (M3)
of the PSPC observation is marked with (x) for the soft band maximum, 
(+) for the hard band maximum and (*) for the broad band maximum. 
The corresponding marginal maximum in the
HRI observation is marked with (o).
}
\end{figure*}

\begin{figure*} \label{fig-sed}
\psfig{figure=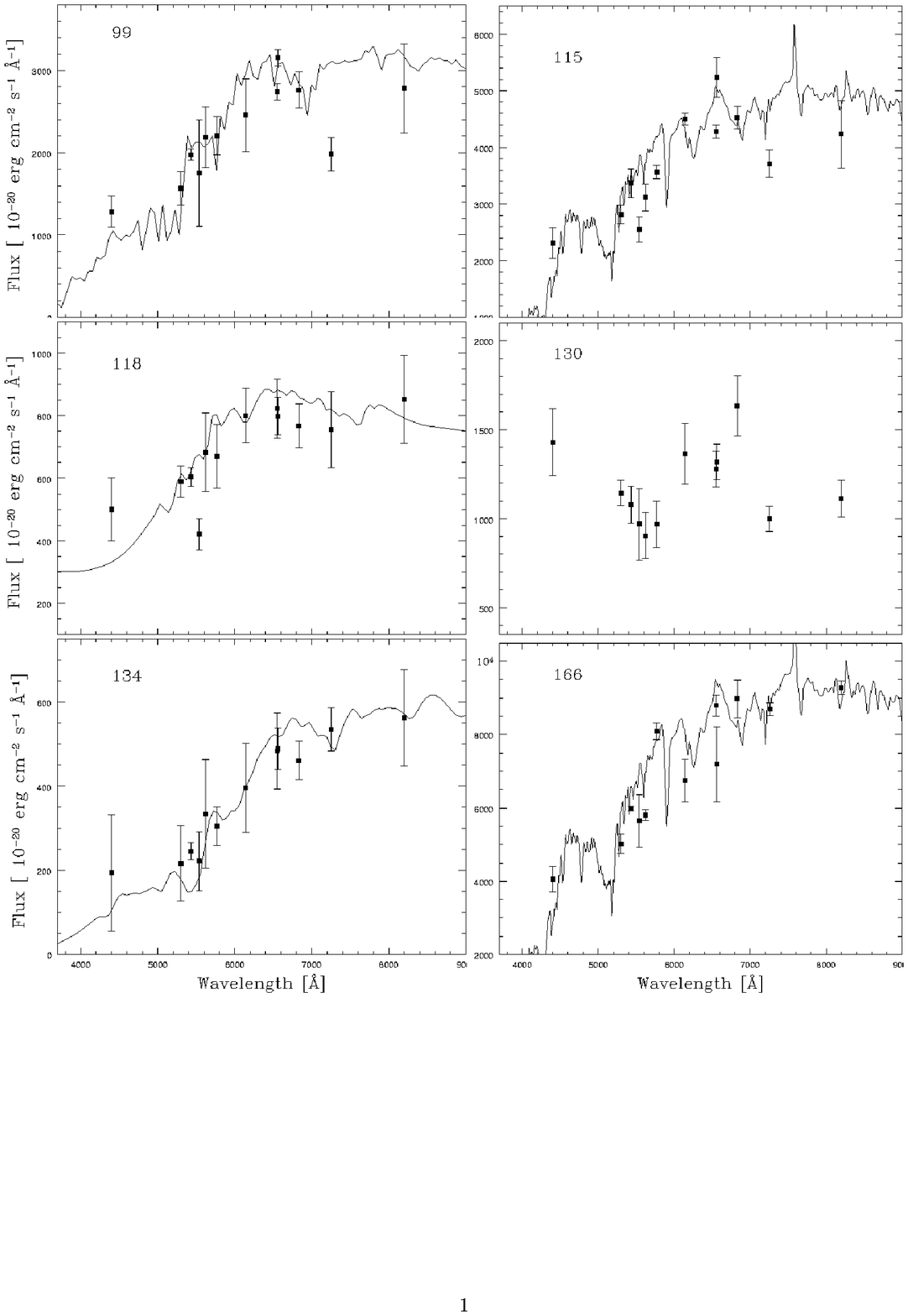,width=17.0cm,clip=} 
\caption[]{Spectral energy distribution for various possible optical
counterparts of RXJ0943.0+4701 (numbers as in Fig. 8). 
All spectral energy distributions -- except for \#130
-- are overlaid with template spectra of probable
identifications. The candidate \#130 could not be uniquely identified. 
Source 
\#99 is overlaid with a spectrum of an Sa galaxy at z=0.33, 
\#115 with a spectrum of a K7V star,
\#118 with an Sb galaxy spectrum at z=0.41, 
\#134 with a spectrum of an elliptical galaxy at z=0.41, and 
\#166 with a M0V star spectrum.  
}
\end{figure*}

The two stars, objects \#166 and \#115, have clearly very red spectra
typical of cool stars. 
A convolution of K- and M-type stars spectra  
 (Gunn \& Stryker  1983) with our filter setup 
shows that both stars can be classified as K6-K8 or  M0-M2   stars.
A possible H$\alpha$ emission would not be visible in the
SEDs because the corresponding filter at 6560 \AA\ has a FWHM of
 170 \AA\ and therefore only a strong emission line
(EW$>$ 30 \AA) could be detected. But it is very rare that these stars have
such is a strong emission line.

For stars of this type an X-ray variability of a factor of ten is 
highly unlikely, unless the star is observed during a flare. 
But a flare can be excluded from the lightcurve, which shows 
approximately constant emission over 3 days 
(see Fig. 7). 
Furthermore, the ratio of X-ray flux to bolometric optical flux would 
be too high for K or M stars. For the brighter star \#166 the flux ratio
$$f_{X,RXJ0943.0+4701} (0.1-2.4keV) / f_{opt,\#166}(bol) 
	\approx 5\times10^{-2},$$
while a value of $10^{-3}$ is the upper limit for K or M stars
	(J. Schmitt, private communication).
Therefore we can rule out the stars as origin of the X-ray emission
of RXJ0943.0\-+4701.

It is also not likely that a cataclysmic variable causes the X-ray
emission because a hardness ratio close to $-1$ (and not the observed
$-0.08$) would be expected.

None of the five galaxies shows the SED or the emission lines 
typical of an AGN. They look indeed like normal spiral galaxies,
either in the foreground or possibly even cluster members.
But we cannot exclude that one of the galaxies changed from an active 
state to a quiescent state within the year that passed between the 
PSPC and the optical observation.
Even for an AGN a variability larger than a factor of 10 is
quite exceptional, but an AGN would be a possible identification.

As seen in the previous section,
a spectral fit of the RXJ0943.0+4701 region with a
power law yields a photon index between 2.2 and 2.9
(with large errors because of
the few photons and contamination from the cluster emission). These
numbers are typical photon indices for AGNs
(e.g. Boller et al. 1996).

Assuming that a cluster member galaxy ($z=0.41$) or the galaxy 
\#99 at a redshift of  about $z=0.33$ were the X-ray emitting AGN we calculate
the luminosity in the ROSAT band.
With a photon index of 2.4 and a Galactic $n_H$ we find a luminosity
log[$L(0.2$keV$)$] = 26.5 
	(corresponding to $L_X$(0.1-2.4keV)$ = 
	4.1\times10^{43}$ erg/s for $z=0.41$) 
or 
log[$L(0.2$keV$)$] = 26.3
	(equivalent to $L_X$(0.1-2.4keV)$ = 
	2.5\times10^{43}$ erg/s for $z=0.33$).
Both numbers are in the standard range for AGNs (Boller et al. 1996).

The values for $f_{X,RXJ0943.0+4701} (0.1-2.4keV) / f_{opt}$ are for all the
galaxies in the typical range for AGNs (Maccacaro et al. 1988).
So an AGN is a viable possibility for the source RXJ0943.0+4701.

The most interesting candidate for the optical identification of
the strongly variable X-ray source RXJ0943.0\-+4701, however, is the blue
compact object \#130. 
With the low-resolution spectra provided by our SED it is not
possible to estimate the redshift since only one possible 
emission line is visible.
However, the spectrum is clearly very blue and the extend of the object
is only slightly larger than the point spread function. 
Therefore the candidate \#130 could be consistent with a quasar/AGN. 
For a true identification we have to wait for a better spectrum, though.
The spectral fits to the RXJ0943.0+4701 X-ray emission 
(see previous section) do not provide strong constraints, 
but these results are certainly not in contradiction with  
RXJ0943.0+4701 being an AGN.
So combining all these different lines of evidence,
our conclusion is that the strongly variable X-ray source RXJ0943.0+4701 is 
most likely an AGN.

\subsection{The quasar at z=2.055} \label{subsec-quasar}

In order to determine the X-ray brightness of the quasar
at redshift z=2.055 (Dressler et al. 1993)
we use the same method as for RXJ0943.0+4701: 
a smoothed model of the point spread function (in this
case the point spread function of the HRI) is subtracted from the
image in Fig. 1. 
We find that the quasar has about 30 counts above the cluster emission 
in the HRI observation. 
This corresponds to a count rate of $7\times10^{-4}$
counts/s. 
Assuming a Galactic hydrogen column density of
$1.27\times10^{20}$cm$^{-2}$ (Dickey \& Lockman 1990) 
and a power law spectrum with photon index = 2.3, we find a
flux $f(0.2$keV$) = 2.7\times 10^{-31}$ erg/cm$^2$/s/Hz and 
a luminosity of log[$L(0.2$keV$)] = 28.1$ 
(corresponding to $L_X$(0.1-2.4keV)$ = 1.4\times 10^{45}$erg/s), 
which places the quasar among the most 
X-ray luminous quasars (La Franca et al. 1995; Green et al. 1995).
However, since the quasar it seen through a galaxy cluster, it
is unavoidable that the gravitational lens effect of the cluster
magnifies the flux of the quasar. Therefore its true X-ray luminosity
is probably somewhat lower.

A comparison of the HRI and the PSPC observation 
(Fig. 6a and b) might suggest a possible X-ray variability of 
the quasar, because  
in the PSPC observation M1 is slightly brighter than M2 and 
it is the other
way round in the HRI image (the latter is smoothed
such that the resolutions of both images are equivalent). But the
number of photons in the two maxima are not significantly
different. As these small differences are within the statistical
fluctuations there is no indication for a brightening
of the quasar between the two observations.

\section{Discussion and Conclusions}\label{sec-dis}

The main results from the X-ray analysis of CL 0939+4713 are
summarized in Table 4. It is quite surprising for such a rich cluster to
have an X-ray luminosity typical for a less rich cluster;
e.g. 
its X-ray luminosity is smaller than the mean luminosity of the 
richness class 4+5 clusters in Soltan \& Henry (1983) and 
CL 0939+4713 is probably richer than these clusters (Dressler et al.
1994). Of course there is a large range of luminosities within each 
richness class. Burg et al. (1994) find a factor of 25 between the 
fifth and 95$^{th}$ percentiles, although there seems to be a
tendency that the range is becoming smaller when going to 
higher richness classes. As 
Burg et al. did the analysis only up to
richness class 2, it is hard to quantify how unusual
the X-ray luminosity of CL 0939+4713 is.

One might try to
explain the low X-ray luminosity 
by the non-equilibrium configuration of the cluster,
which is obvious from the pronounced substructure visible in the
ROSAT/HRI image. Evidently, there are still
separate units in this cluster which are still in the process of merging.
After the final merging the cluster will have probably a
higher central density and -- as the luminosity is proportional to
the square of the density -- also a higher luminosity in this stage. 
But as this effect is not expected to be large (Schindler \& M\"uller
1993) it cannot explain the relatively low luminosity of this
obviously pre-merger cluster.

\begin{table}[htbp]
\begin{center}
\begin{tabular}{|c|c|c|}
\hline
&  \multispan2 \vrule \\
countrate HRI [counts/s]     &\multispan2\hfill 0.023\hfill\vrule \\ 
$L_X$(0.1-2.4keV) [erg/s]&\multispan2\hfill$6.4_{-0.3}^{+0.7}\times 10^{44}$\hfill\vrule  \\
$L_X$(bol) [erg/s]       &\multispan2\hfill$1.6_{-0.3}^{+0.5}\times 10^{45}$\hfill\vrule  \\
temperature [keV]        &\multispan2\hfill$7.6_{-1.6}^{+2.8}$ \hfill\vrule \\
metallicity [solar]&\multispan2\hfill$0.22_{-0.22}^{+0.24}$\hfill\vrule\\
$n_H$ [cm$^{-2}$]&\multispan2\hfill $0.9_{-0.4}^{+0.5}\times10^{20}$\hfill\vrule  \\
& \multispan2 \vrule \\
\hline
& &   \\
subcluster                & M1 & M2 \\
& &   \\
\hline
& &   \\
$M_{gas}$($r\le 265$ kpc)[$\msol$]& $4.1\times 10^{12}$ 
                                  & $4.2\times 10^{12}$  \\ 
$M_{tot}$($r\le 265$ kpc)[$\msol$]& $6.0_{-1.6}^{+2.3}\times 10^{13}$
                                  & $7.4_{-2.0}^{+2.9}\times 10^{13}$  \\
gas mass fraction                 & $7\pm2\%$ & $6\pm2\%$ \\
& &   \\
\hline
\end{tabular}
\end{center}
\caption{Summary of the X-ray properties of CL 0939+4713, and its two
subclusters M1 and M2.
}
\end{table}

The temperature could be determined here without the contamination of
the variable source RXJ0943.0+4701 to $T=7.6_{-1.6}^{+2.8}$ keV. 
The luminosity-temperature relation  
for high redshift clusters 
(e.g. Mushotzky \& Scharf 1997; Tsuru et al. 1997) 
predicts a temperature around 6 keV, i.e. the
temperature of CL 0939+4713 is somewhat higher than expected from its 
luminosity, but still consistent within the scatter of the 
luminosity-temperature relation.
From the point of view of
the velocity dispersion the temperature is consistent within the 
scatter of the velocity-temperature relation (Edge \& Stewart 1991;
Lubin \& Bahcall 1993; Bird et al. 1995), but again slightly on the high 
temperature  side.
Taking into account the substructure in the cluster one can
explain such a behaviour by a merger roughly 
perpendicular to the line-of-sight.
The increase of temperature (and of luminosity) is visible from all
directions while the increase in the velocity dispersion is very small
when it is observed from a direction perpendicular to the collision axis.

It seems puzzling that a cluster with such a high galaxy content has
not a high metallicity. The metallicity of $0.22_{-0.22}^{+0.24}$ is
in the normal range compared with other clusters with this temperature (and
less galaxies) which have
Fe abundances around 0.3 (Fukazawa et al. 1996, Mushotzky \&
Loewenstein 1997). The ratio of iron mass to light makes the
difference even more evident. 
The value found in CL 0939+4713 of
$M_{Fe}/L_B \approx 5\times10^{-3}\msol/\lsol$ is smaller than the one
expected for
rich clusters of 0.01-0.02 (Renzini et al. 1993), although one must 
keep in mind the errors are so large
that there is some overlap and, in addition, the observed dispersion 
in metallicity is about a factor of two (Mushotzky \& Loewenstein 1997). 

In order to find an explanation for 
this difference we have a closer look into the galaxy
population of the cluster, having in mind the conclusions of 
Arnaud et al. (1992) that only ellipticals and lenticulars contribute to
the metal enrichment of in the intra-cluster medium.
Andreon et al. (1997) found from HST images that CL 0939+4713 is
overabundant in spirals. They compared the percentage of
ellipticals and lenticulars with the one in the Coma cluster. While
they find about 50\% E's and S0's of identified galaxies in CL 0939+4713,
80\% of the Coma cluster galaxies are E's and S0's. Furthermore, we 
can compare it
with another cluster which is not so well virialised as the Coma
cluster but has also pronounced substructure like CL 0939+4713: the
Virgo cluster. This cluster has a
percentage of ellipticals and lenticulars of 81\% (Binggeli et
al. 1985) --  also much larger than in CL 0939+4713.
These are all results which are based on a morphological 
classification. Using a spectral analysis one finds a somewhat different
result for CL 0939+4713 (Belloni et al. 1995): ellipticals 67\%,
Spirals and Im 15\%, E+A galaxies 19\%.  
The high fraction of  E+A
galaxies would be classified in the morphological analysis as
ellipticals. But as they have a preceding starburst phase, 
they are in terms of the iron enrichment more similar to
spirals. Therefore one should not count the E+A galaxies as
ellipticals, so that the elliptical fraction is just 67\%. This number
is larger than the one from the morphological study but still smaller
than other elliptical percentages. Summarizing, the low elliptical
content together with the low iron mass to light ratio confirms the
correlation of elliptical and iron content found by Arnaud et
al. (1992).  

The gas mass fraction in this cluster of 6 or 7\% 
is lower than in other clusters, where 10-30\% are observed 
(B\"ohringer 1995). The small values can be either due to an error
in the determination in the total mass, as the total mass for
non-relaxed clusters is general difficult to determine (Ota et al. 
1998; Allen 1997). A major problem can be projection effects of 
subclusters like e.g. in CL 0500-24 (Schindler \& Wambsganss 1997). 
Cen (1997) found that these projection effects may be
able to account for all discrepancies found between X-ray masses and 
lensing masses. 

On the other hand the consistency with the 
luminosity-temperature relation implies together with the small gas 
mass fraction a compact structure of the subclusters, which is
found in the subclusters around M1 and M2 (see small core radii in
Table 1). For a standard-shaped isothermal cluster the temperature 
has a fixed relation with the total mass, as well as the X-ray 
luminosity with the gas mass. To get a low gas mass fraction a 
more compact gas distribution than the average one is necessary, which 
produces the same X-ray luminosity with a smaller amount of
gas. As this is the case in CL 0939+4713 everything gives a consistent    
picture. The mass to light ratio of $M/L_B
 \approx 
350\msol/\lsol$ is also rougly in the standard range.
This implies that there is not necessarily an error
in the mass estimate and the low gas mass fraction can be real. Such 
variations in the baryonic fractions would have implications 
for cosmology. They would be explainable either by  
dynamical processes operating 
differently on baryonic and dark matter or by pre-inflationary 
baryon-to-total mass fluctuations which are preserve d 
into the post-inflationary epoch (see Evrard 1997).

This new X-ray analysis of ROSAT/HRI and ASCA data confirms that 
CL 0939+4713 is a young cluster. It is a cluster where
mergers on different levels are taking place, which is obvious from the
complex subcluster structure. Also other properties point in this
direction: for its richness the cluster has a relatively low
luminosity, low gas mass, low total mass and low iron mass, which
are all indicators for young systems.

The variable source RXJ0943.0+4701 at a distance of
about 1.5 arcmin from the cluster centre
was discovered by comparing the ROSAT/PSPC and the ROSAT/HRI image.
Its luminosity decreased
by at least a factor ten within the five years between the two
ROSAT observations. Only a small fraction of X-ray sources has
such large variations in the flux (Voges \& Boller 1997). 

To identify RXJ0943.0+4701 we use the
X-ray properties (X-ray spectrum, short term time variability) in
combination with optical properties taken from observations with
various filters (SED, ratio of optical to X-ray
flux). This comparison rules out stars and cataclysmic variables as
origin of the X-ray emission of RXJ0943.0+4701, so that the most
likely origin would be an AGN. 

So far only very few AGNs have
been found with large X-ray variabilities and these strongly
variable AGNs turned out to be mainly narrow-line Seyfert 1 galaxies
(e.g. Piro et al. 1988; Grupe et al. 1995, Otani et al. 1996, Boller
et al. 1997). It is not clear yet what
physical processes are responsible for the strong X-ray variability.
Possible explanations (as discussed in Boller et
al. 1997) could be strong relativistic effects and partial covering by
occulting structures in an accretion disc. Optical follow-up
observations will give us detailed information on this interesting
object.

\begin{acknowledgements}
It is a pleasure to thank Doris Neumann for providing the
code to determine the significance of substructure, 
Carlo Izzo for his most helpful EXSAS support,
Thomas Boller for assistance with the time analysis, and Stella Seitz
for determining the lensing mass for us on the complicated overlapping
area. We
thank D. Grupe, J.H.M.M. Schmitt and H.C. Thomas for helpful
discussions about variable sources.
The referee Richard Mushotzky is thanked for valuable suggestions.
S.~S. acknowledges 
financial support by the Verbundforschung.
\end{acknowledgements}
%
%

\end{document}